%% file: iclr2025_conference.tex
\documentclass{article} 
\usepackage{iclr2025_conference,times}

\input{math_commands.tex}

\usepackage{hyperref}
\usepackage{url}
\usepackage{wrapfig}
\usepackage{graphicx}
\usepackage{booktabs}

\usepackage{fontawesome5}
\usepackage{booktabs}
\usepackage[table]{xcolor}
\usepackage{makecell}
\usepackage{tabularx}

\definecolor{adprcolor}{HTML}{4198AC}
\definecolor{tprcolor}{HTML}{ED8D5A}

\definecolor{tablegray}{gray}{0.965}
\definecolor{tablehead}{gray}{0.92}
\definecolor{bestblue}{RGB}{225,237,247}
\definecolor{focusorange}{RGB}{252,236,220}
\definecolor{modelband}{RGB}{244,241,235}

\newcommand{\best}[1]{\cellcolor{bestblue}\textbf{#1}}
\newcommand{\focus}[1]{\cellcolor{focusorange}\textbf{#1}}

\newcolumntype{L}{>{\raggedright\arraybackslash}X}
\newcolumntype{Y}{>{\centering\arraybackslash}X}

\title{AsynCodeBench: Benchmarking\\ Collaboration of Asynchronous Multi-Agent Systems in Software Engineering} 

\author{
Kaituo Zhang \\
University of Houston
\And
Zhen Xiong \\
New York University
\And
Zhimeng Jiang \\
Texas A\&M University
\And
Mingyu Zhong \\
University of Houston
\And
Zhouyuan Yuan \\
UCSD
\And
Zhecheng Li \\
UCSD
\And
Bowen Lin \\
University of Houston
\And
Chia-Yuan Chang \\
Texas A\&M University
\And
Mingzhi Hu \\
Worcester Polytechnic Institute
\And
Huazheng Wang \\
Oregon State University
\And
Ying Lin \\
University of Houston
}

\iclrfinalcopy 
\begin{document}

\maketitle

\begin{abstract}
Multi-agent coding has emerged as an increasingly active direction in software engineering, where complex development tasks are decomposed across multiple specialized agents working on different parts of the problem. Despite the shift from individual problem solving to distributed collaboration, multi-agent systems still lack a direct measure of collaboration and are largely evaluated through task-level outcomes inherited from single-agent coding, conflating individual coding capability with cross-agent coordination. We introduce \textbf{AsynCodeBench}, a dependency-centric benchmark for asynchronous multi-agent software engineering that represents each task with an explicit dependency graph and executable Dependency Checkers. Through this dependency-tracking process, we propose two complementary measures: \textit{\underline{A}synchronous \underline{D}ependency \underline{P}ass \underline{R}ate} (ADPR), which measures how many cross-agent dependencies are ultimately satisfied, and \textit{\underline{D}ependency \underline{R}esolution \underline{S}tep (DRS)}, which measures when each dependency first becomes satisfied during execution. AsynCodeBench comprises 19 tasks from real-world repositories, exposing 52 directed dependencies as explicit units for evaluating cross-agent collaboration. Experiments across model families, scales, and generations reveal a clear gap between coding and collaboration capability: improvements in coding performance do not necessarily translate into stronger collaboration, and task-level metrics can diverge substantially from dependency-level collaboration measures. Dependency-trajectory analysis further reveals that successful coordination often emerges not gradually, but through concentrated bursts in which many dependencies become resolved over a short portion of the execution trajectory, a pattern we term a hopping window. These findings position dependency dynamics as a useful lens for understanding and interpreting collaboration behavior in asynchronous multi-agent software engineering.
\end{abstract}

\vspace{-0.3em}
\begin{center}
\small
\faGithub\ 
\href{https://github.com/KaituoZhang/AsynCodeBench/tree/main}
{\texttt{github.com/KaituoZhang/AsynCodeBench}}
\qquad
\faLink\ 
\href{https://AsynCodeBench.org}
{\texttt{AsynCodeBench.org}}
\end{center}
\vspace{-0.4em}



\input{content/Introduction}

\input{content/Related_Work}

\input{content/Method}

\input{content/Experiment}

\input{content/Discussion}

\bibliography{iclr2025_conference}
\bibliographystyle{iclr2025_conference}

\appendix

\input{content/Appendix}

\end{document}

%% file: math_commands.tex
\usepackage{amsmath,amsfonts,bm}

\def\eqref#1{equation~\ref{#1}}

\def\1{\bm{1}}

\DeclareMathAlphabet{\mathsfit}{\encodingdefault}{\sfdefault}{m}{sl}
\SetMathAlphabet{\mathsfit}{bold}{\encodingdefault}{\sfdefault}{bx}{n}



%% file: content/Introduction.tex
\section{Introduction}

Large language model agents have rapidly expanded from code generation to repository-level software engineering, where they can modify files, execute tests, and iteratively repair implementations~\citep{jimenez2024swebench,yang2024sweagent}. Multi-agent software development has consequently emerged as an active direction, with systems such as ChatDev, MetaGPT, RTADev, CAID, and CodeTeam distributing planning, implementation, testing, and coordination across specialized agents~\citep{qian2024chatdev,hong2024metagpt,liu2025rtadev,wang2026codeteam,geng2026effectivestrategiesasynchronoussoftware}. Recent work has also begun to study interaction and coordination more explicitly, including recovery from out-of-sync states and corrective interaction during coding~\citep{guo2025syncmind,wu2026swetogether}. Yet multi-agent software engineering introduces a capability beyond solving the task itself: agents must coordinate their independently produced work. Strong performance in single-agent coding therefore does not necessarily imply effective collaboration in a multi-agent setting.

Evaluation, however, remains largely focused on final task outcomes. Repository-level benchmarks such as SWE-bench assess the final implementation through executable tests~\citep{jimenez2024swebench}, while multi-agent coding systems often report gains using the same end-task metrics over single-agent baselines~\citep{geng2026effectivestrategiesasynchronoussoftware,wang2026codeteam}. These comparisons show whether multi-agent execution performs better, but do not directly measure collaboration: final scores conflate individual coding ability with cross-agent coordination and may hide unresolved dependencies across independently implemented components.

We therefore use cross-component software dependencies as a lens for collaboration and introduce \textbf{AsynCodeBench}, a dependency-centric benchmark for asynchronous multi-agent software engineering. Unlike existing evaluations that characterize collaboration through final task outcomes or isolated interaction behaviors, AsynCodeBench makes cross-agent dependency resolution observable throughout execution. We introduce Asynchronous Dependency Pass Rate (ADPR) to capture dependency resolution and Dependency Resolution Step (DRS) to capture resolution timing, enabling collaboration to be evaluated both by its outcome and by how it develops over time.

Our study provides a dependency-centric view of collaboration in multi-agent software engineering, separating collaboration capability from conventional coding performance and making its evolution observable throughout execution. Using AsynCodeBench, we find that multi-agent execution is not uniformly stronger than single-agent execution, and that within the evaluated Qwen family, stronger single-agent coding performance does not improve all aspects of collaboration. Dependency trajectories further reveal a recurring \textit{hopping} phenomenon, where substantial dependency resolution is often established through concentrated phases whose timing varies across models and execution protocols. Finally, dependency-aware evaluation can lead to different quality--cost assessments from conventional test-based metrics. These findings highlight collaboration as a distinct capability that cannot be fully characterized by final coding outcomes alone.

%% file: content/Related_Work.tex
\section{Related Work}
\label{sec:related_work}

\subsection{Repository-Level Coding Agents and Benchmarks}

Repository-level coding benchmarks have emerged to bridge the gap between closed-form code generation and real software engineering, where solving a task requires reasoning over an existing codebase rather than completing an isolated function~\citep{ding2023crosscodeeval,liu2024repobench,jimenez2024swebench}. This setting preserves challenges that are largely abstracted away by function-level evaluation, including cross-file reasoning, software dependencies, execution feedback, and consistency across interacting changes~\citep{gautam2025refactorbench,li2025feabench,deng2025swebenchpro,le2026sweevobenchmarkingcodingagents,ding2025nl2repobench}. To retain these challenges, benchmark construction has increasingly relied on real repositories and development artifacts together with executable validation, preserving much of the surrounding software context encountered in practice~\citep{jimenez2024swebench,li2025feabench,pan2025swegym,miserendino2025swelancer,badertdinov2025swerebench,vergopoulos2025automated}. This repository-grounded paradigm has since expanded across software ecosystems and toward increasingly broad and long-horizon development tasks, ranging from feature and library implementation to software evolution and repository-scale transformation~\citep{yang2024swebenchmultimodal,zan2025multiswebench,rashid2025swepolybench,zhao2024commit0librarygenerationscratch,yang2025swesmith,badertdinov2026swerebenchv2,le2026sweevobenchmarkingcodingagents,hong2026swerefactorbenchcoding}. Yet as the tasks themselves have become more realistic and extended over time, their evaluation still largely reduces execution to final task completion, leaving intermediate progress and failure dynamics comparatively under-characterized.

\subsection{Multi-Agent Software Engineering}

Multi-agent software engineering has rapidly expanded from role-based workflows toward repository-scale collaboration, with increasingly sophisticated mechanisms for specialization, delegation, integration, and concurrent execution~\citep{qian2024chatdev,hong2024metagpt,liu2025rtadev,wang2026codeteam,oliveira2026developing,costa2026agentspawn,kumar2026agentforge}. Despite this shift, evaluation still largely inherits a single-agent perspective, judging multi-agent systems by end-task performance or gains over single-agent baselines rather than measuring collaboration itself~\citep{khatua2026cooperbench,donato2026rolebased,destefanis2026whenagentscoordinate}. Recent work has begun to expose this gap through failure analysis, out-of-sync recovery, role-separated coordination, distributed communication, and interactive coding~\citep{cemri2025mast,guo2025syncmind,kim2026teambench,zhang2026silobench,wu2026swetogether}. AsynCodeBench goes further by tracing the dependencies that must be resolved across independently implemented components, revealing failed handoffs and coordination dynamics that can remain hidden behind final task performance.

%% file: content/Method.tex
\section{Benchmark Construction}

\subsection{Problem Definition}

Multi-agent software engineering often decomposes a coding task across multiple agents, each responsible for a different implementation unit. We refer to collaboration as \textbf{asynchronous} when these agents proceed independently without necessarily waiting for dependent agents to complete or propagate their updates. When dependencies exist across their subproblems, this execution 
\begin{wrapfigure}{r}{0.5\linewidth}
    \centering
    \includegraphics[width=\linewidth]{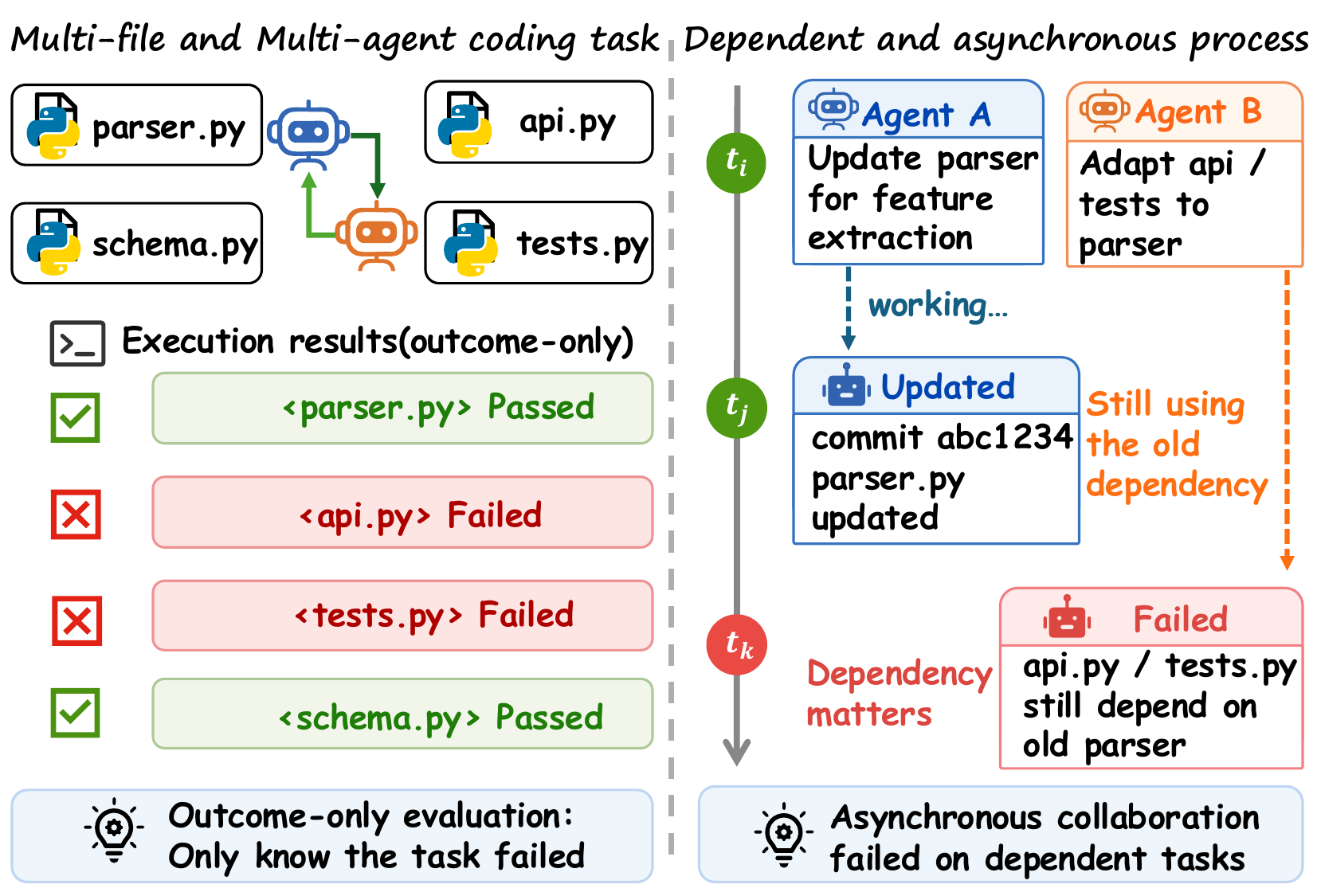}
    \caption{A small illustration of the existing gap.}
    \label{fig:async-problem}
\end{wrapfigure}pattern can create coordination failures: a downstream agent may continue working with incomplete or outdated upstream behavior, causing the independently produced implementations to become inconsistent when they are integrated. We refer to failures arising from this interaction between asynchronous execution and software dependencies as the \textbf{asynchronous coordination problem} studied in AsynCodeBench. A key difficulty is that these failures are often hidden by traditional outcome-only evaluation, since dependencies across agents' subproblems are typically implicit and not explicitly tracked. As illustrated in Figure~\ref{fig:async-problem}, the evaluator may identify failed components without revealing the stale dependency between agents that produced them. This makes it difficult to determine whether the failure reflects limitations in individual coding capability or in multi-agent collaboration.

The key question is therefore whether dependencies between agents' implementations are correctly maintained during collaboration, rather than only whether the final repository succeeds. In AsynCodeBench, a dependency exists when the correctness of code assigned to one agent depends on code implemented by another, such as a required function, interface, data structure, or shared state. We make these dependencies explicit through task-specific dependency graphs, evaluate them with executable Dependency Checkers, and track them across integration checkpoints to determine whether and when they are resolved.

\subsection{Dependency Graph and Executable Checkers}
\label{sec:dependency-checkers}

We operationalize asynchronous coordination through four benchmark components,
\[
\mathcal{B}=(\mathcal{R},\mathcal{G},\mathcal{C},\mathcal{E}),
\]
where \(\mathcal{B}\) denotes AsynCodeBench, \(\mathcal{R}\) the repository and target coding task, \(\mathcal{G}\) the task-specific dependency graph, \(\mathcal{C}\) the executable Dependency Checkers defined over the dependencies in \(\mathcal{G}\), and \(\mathcal{E}\) the benchmark evaluation.

Each task is represented by a directed dependency graph
\[
\mathcal{G}=(\mathcal{V},\mathcal{A}),
\]
where \(\mathcal{V}\) is the set of implementation units and \(\mathcal{A}\subseteq\mathcal{V}\times\mathcal{V}\) is the set of directed dependency arcs. Each \(d\in\mathcal{A}\) connects an upstream producer to a downstream consumer whose correctness depends on behavior provided by the producer. The graph is defined before model evaluation. Appendix~\ref{subsec:Depedency_Task} details the dependency construction procedure and supporting repository evidence.

For each dependency \(d\in\mathcal{A}\), we pre-register three groups of executable Boolean checkers,
\[
\mathcal{C}_d=
\left(\mathcal{C}_d^{U},\mathcal{C}_d^{D},\mathcal{C}_d^{I}\right),
\qquad
\mathcal{C}=\{\mathcal{C}_d\}_{d\in\mathcal{A}}.
\]
Each constituent checker \(c\) returns a binary pass/fail outcome on a repository workspace \(W\), with \(c(W)=1\) indicating that the tested condition is satisfied. The upstream group \(\mathcal{C}_d^{U}\) checks whether the producer provides the required behavior, the downstream group \(\mathcal{C}_d^{D}\) checks whether the consumer correctly uses or preserves it, and the integrated group \(\mathcal{C}_d^{I}\) checks whether the producer--consumer dependency holds after integration. For \(X\in\{U,D,I\}\), the group-level outcome is
\[
C_d^{X}(W)=\mathbf{1}\!\left[\forall c\in\mathcal{C}_d^{X},\,c(W)=1\right].
\]
The upstream and downstream outcomes provide diagnostic evidence about the two sides of a dependency, whereas \(C_d^{I}\) determines whether the dependency itself is satisfied in the integrated workspace.

Let \(W_t\) denote the retained integrated workspace at checkpoint \(t\), \(t=1,\ldots,T\). We define
\[
I_{d,t}=C_d^{I}(W_t),
\qquad
\mathbf{I}_d=(I_{d,1},\ldots,I_{d,T}),
\]
where \(I_{d,t}\in\{0,1\}\) is the satisfaction state of dependency \(d\) at checkpoint \(t\). Because the same dependency is evaluated over an ordered sequence of integrated repository states, \(\mathbf{I}_d\) records its resolution trajectory over execution. It therefore distinguishes dependencies that never resolve, resolve only after several integration steps, or resolve and later regress---information that is lost when only the final workspace is evaluated. Appendix~\ref{app:checkpoint-construction} details checkpoint construction and indexing.

\subsection{Evaluation Metrics}
\label{sec:metrics}

\paragraph{Test Pass Rate.}
Unlike a binary task success measure, Test Pass Rate quantifies the fraction of evaluation points obtained from the repository-level test suite:
\[
\operatorname{TestPass}=\frac{P}{N},
\]
where \(P\) is the number of passed test points and \(N\) is the total number of test points for the current task. It provides proportional credit for partially correct implementations, but does not directly assess whether the benchmark-defined cross-component dependencies are satisfied.

\paragraph{Asynchronous Dependency Pass Rate.}
Using the pre-registered dependency graph and executable checker states defined above, ADPR measures the fraction of dependencies satisfied in the final integrated workspace:
\[
\operatorname{ADPR}
=
\frac{1}{|\mathcal{A}|}
\sum_{d\in\mathcal{A}} I_{d,T}.
\]
ADPR therefore provides a dependency-level measure of how successfully the cross-component dependencies identified in AsynCodeBench are resolved after integration.

\paragraph{Dependency Resolution Step.}
Based on the ordered integrated-checker outcomes defined above, DRS measures the first checkpoint at which a dependency passes its integrated checker:
\[
\operatorname{DRS}_d
=
\min\{t:I_{d,t}=1\},
\]
with \(\operatorname{DRS}_d=T+1\) when dependency \(d\) never passes during execution. Since the number of retained checkpoints varies across runs, we normalize the resolution step as
\[
\widetilde{\operatorname{DRS}}_d
=
\frac{\operatorname{DRS}_d}{T+1},
\]
where smaller values indicate earlier resolution and \(1\) denotes non-resolution. By using the ordered dependency trajectory rather than only its final state, DRS distinguishes dependencies that resolve early from those that become valid only after later integration steps, providing a temporal view of dependency resolution that final ADPR alone cannot capture. DRS is defined over logical integration checkpoints rather than wall-clock time. Since \textsc{Single} exposes only the final workspace, it does not provide a dependency trajectory and is therefore excluded from DRS.

\subsection{Task Curation and Human Validation}

We curate public repository tasks whose implementations naturally span behaviorally coupled files or modules, preserving existing software boundaries rather than introducing artificial code partitions. AsynCodeBench contains 19 tasks from 16 repositories, covering framework and library engineering, systems and storage, compiler/IR engineering, network protocols, and security. Each task is packaged with a fixed repository revision, an answer-free task instruction, scoped writable paths, and a task-specific \texttt{pytest} evaluator. All execution protocols use the same task package, OpenHands~\citep{wang2025openhandsopenplatformai} harness, and task-specific Docker environment.

To ensure that the dependency structure reflects genuine repository constraints, all dependency graphs undergo independent human validation. Three reviewers verify the validity and direction of each dependency together with its supporting executable evidence, and only validated dependencies are retained. The complete task inventory, packaging details, and dependency-validation protocol are reported in Appendix~\ref{app:task-inventory}, with human-review details in Appendix~\ref{app:human_review_details}.

\section{Evaluation Protocols}
\label{sec:protocols}

We organize the five execution protocols as a progression from independent full-task solving to increasingly structured collaboration. \textsc{Single} assigns the entire task to one agent. \textsc{Serial} introduces specialist decomposition while preserving producer-to-consumer handoffs, whereas \textsc{Async-Private} keeps the same decomposition but allows specialists to proceed concurrently in private workspaces. The two manager variants then add an explicit coordination layer, first without write access and then with bounded repair.

The contrast between \textsc{Serial} and \textsc{Async-Private} isolates the effect of asynchronous execution under the same specialist decomposition. In \textsc{Serial}, each downstream specialist begins only after the corresponding upstream artifact has been integrated; in \textsc{Async-Private}, specialists work from the same initial repository state and cannot observe one another's in-flight changes. This creates the setting in which unresolved cross-agent dependencies can emerge.

The manager variants test whether these failures can be mitigated through coordination. \textsc{Async-Manager-RO} allows a read-only manager to inspect artifacts, provide feedback, and coordinate integration without modifying production code. \textsc{Async-Manager} further permits bounded repair after specialist integration. This distinction separates coordination through information exchange from direct code intervention. Full protocol details are provided in appendix~\ref{app:protocl_details}.

%% file: content/Experiment.tex
\section{Experiments}
\label{sec:experiments}

\subsection{Experiment Settings}

We evaluate ten open-weight models from the Qwen, DeepSeek, Gemma, Muse, GLM, and Nemotron families, covering dense and mixture-of-experts architectures from 9B dense parameters to 284B total parameters. We use shortened model names. Complete checkpoint identifiers, parameter counts, architectures, and weight precisions are provided in Appendix~\ref{app:model-details}.

Using AsynCodeBench, we systematically investigate multi-agent software engineering across four research questions, progressing from collaboration effectiveness and model capability to dependency-resolution dynamics, and resource costs.

\begin{enumerate}
    \item[\textbf{RQ1}] How does multi-agent collaboration compare with single-agent execution, and what aspects of collaboration are overlooked by conventional task-level metrics?

    \item[\textbf{RQ2}] How do model scale and generation relate to collaborative performance and dependency resolution?

    \item[\textbf{RQ3}] How do execution protocols shape dependency-resolution outcomes and closure dynamics over successive integration checkpoints?

    \item[\textbf{RQ4}] How do token costs vary across execution protocols, and how does the choice of quality metric affect the observed quality--cost trade-offs?

\end{enumerate}

\subsection{Collaboration Gains and Hidden Dependency Failures}
\label{sec:collaboration-gap}

\begin{figure*}[t]
    \centering
    \includegraphics[width=\textwidth]{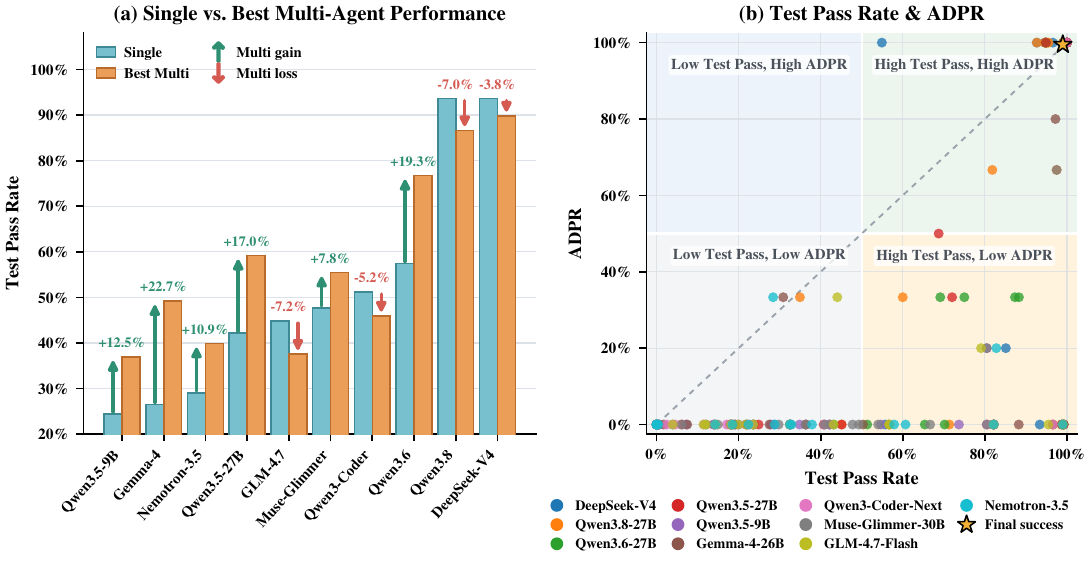}
    \caption{Collaboration gains and hidden dependency failures. (a) For each model, \textsc{Single} is compared with the highest task-macro Test Pass Rate achieved by any of the four multi-agent protocols; arrows show the signed gain. (b) Test Pass Rate versus ADPR for 190 \textsc{Async-Manager-RO} model--task outcomes. The dashed line indicates equality between the two metrics, and the 50\% boundaries provide a visual partition.}
    \label{fig:collaboration-gap}
\end{figure*}

We first ask whether multi-agent execution consistently improves over standalone coding across models. Figure~\ref{fig:collaboration-gap}(a) compares each model's task-macro Test Pass Rate under Single with its best result among the four multi-agent protocols. This Best Multi score is intentionally optimistic: it asks whether any collaborative protocol can improve upon the same model acting alone, without yet attributing the gain to a particular protocol.

The answer varies markedly across models. Six of the ten models improve under their best multi-agent protocol, with gains of 7.8--22.7\%, while four remain worse than \textsc{Single} even after selecting their best collaborative protocol. The contrast is particularly clear for models with high standalone performance: Qwen3.8-27B and DeepSeek-V4 both reach a 93.6\% Test Pass Rate under \textsc{Single}, yet their best multi-agent results decrease to 86.7\% and 89.9\%, respectively. Thus, multi-agent decomposition can provide substantial gains, but these gains are neither universal nor guaranteed to preserve a model's standalone performance.

Task-level progress, however, does not alone reveal whether collaboration succeeds across component boundaries. Figure~\ref{fig:collaboration-gap}(b) examines this distinction under \textsc{Async-Manager-RO} using all 190 model--task outcomes. Test Pass Rate and ADPR are positively but imperfectly associated (Pearson $r=0.617$; Spearman $\rho=0.613$), and their disagreement is strongly asymmetric. Test Pass Rate exceeds ADPR in 136 of 190 outcomes (71.6\%), with a mean Test Pass Rate of 48.0\% compared with only 18.8\% ADPR. More strikingly, 32 outcomes achieve at least 80\% Test Pass Rate while resolving fewer than half of their declared dependencies, including 26 with zero ADPR; we observe no corresponding low-test, high-ADPR cases. Results for the other four protocols are in Appendix~\ref{appendix:test_pass_rate_adpr}.

Overall, multi-agent execution is not uniformly beneficial, and task-level success alone provides an incomplete picture of collaboration. While Test Pass Rate measures broad implementation progress, ADPR directly captures whether dependencies across specialists are successfully resolved, exposing coordination failures that may remain hidden behind high aggregate test performance.

\subsection{Model Capability: Qwen Scale and Generation}
\label{sec:qwen-capability}

\begin{figure*}[t]
    \centering
    \includegraphics[width=\textwidth]{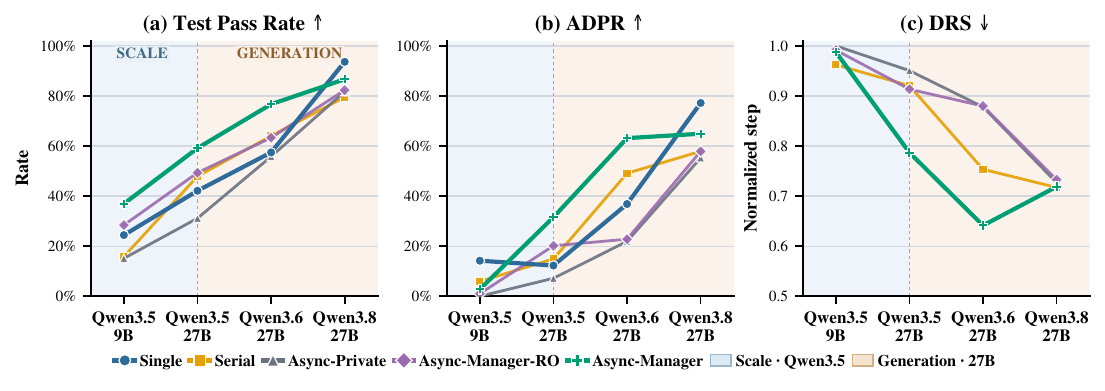}
    \caption{Qwen scale and generation analysis under five matched execution protocols. The left region compares Qwen3.5-9B and Qwen3.5-27B, while the right region compares Qwen3.5, Qwen3.6, and Qwen3.8 at a nominal scale of 27B. Panels report Test Pass Rate, ADPR, and normalized DRS; lower DRS indicates earlier dependency resolution.}
    \label{fig:qwen-scale-generation}
\end{figure*}

The cross-model results above suggest that improvements in standalone coding performance do not necessarily translate into comparable gains under multi-agent execution. To examine this pattern under a more controlled setting, we turn to the Qwen family. We compare Qwen3.5-9B with Qwen3.5-27B to study scaling within the same generation, and Qwen3.5-27B, Qwen3.6-27B, and Qwen3.8-27B to study model evolution at a fixed nominal scale. Figure~\ref{fig:qwen-scale-generation} reports Test Pass Rate, ADPR, and normalized DRS.

Scaling Qwen3.5 from 9B to 27B consistently improves overall test progress, but its effect on dependency resolution is less uniform. Test Pass Rate increases under every protocol, including from 24.4\% to 42.1\% for \textsc{Single} and from 36.9\% to 59.2\% for \textsc{Async-Manager}. ADPR follows a different pattern: \textsc{Single} changes only from 14.2\% to 12.3\%, while \textsc{Async-Manager} increases from 2.8\% to 31.6\%. Thus, additional model capacity clearly improves implementation progress, but the extent to which this improvement carries over to cross-component dependency resolution depends strongly on the execution protocol.

The contrast becomes more pronounced across generations. At 27B, \textsc{Single} improves substantially from Qwen3.5 to Qwen3.6 and Qwen3.8, with Test Pass Rate increasing from 42.1\% to 57.4\% to 93.6\% and ADPR from 12.3\% to 36.8\% to 77.2\%. Multi-agent execution also improves, but not at the same rate. Under \textsc{Async-Manager}, ADPR rises from 31.6\% for Qwen3.5 to 63.2\% for Qwen3.6, but changes only marginally to 64.9\% for Qwen3.8. In other words, the large improvement observed in standalone dependency resolution from Qwen3.6 to Qwen3.8 is not matched by a comparable gain under collaborative execution.

The trajectory analysis reveals a further distinction that final ADPR alone cannot capture. Under \textsc{Async-Manager}, normalized DRS decreases from 0.787 for Qwen3.5 to 0.641 for Qwen3.6, indicating earlier dependency resolution, but rises again to 0.719 for Qwen3.8 despite their similar final ADPR. Two models can therefore reach comparable final dependency coverage while arriving there through different coordination trajectories.

\subsection{Protocol Effects and Dependency Closure Dynamics}
\label{exp:protocol}

The Qwen analysis shows that stronger standalone capability does not uniformly translate into stronger collaboration, and that similar final dependency coverage can arise from different resolution trajectories. We therefore next examine the other side of the interaction: how the execution protocol shapes both the amount and timing of dependency resolution.

\begin{figure*}[t]
    \centering
    \includegraphics[width=\textwidth]{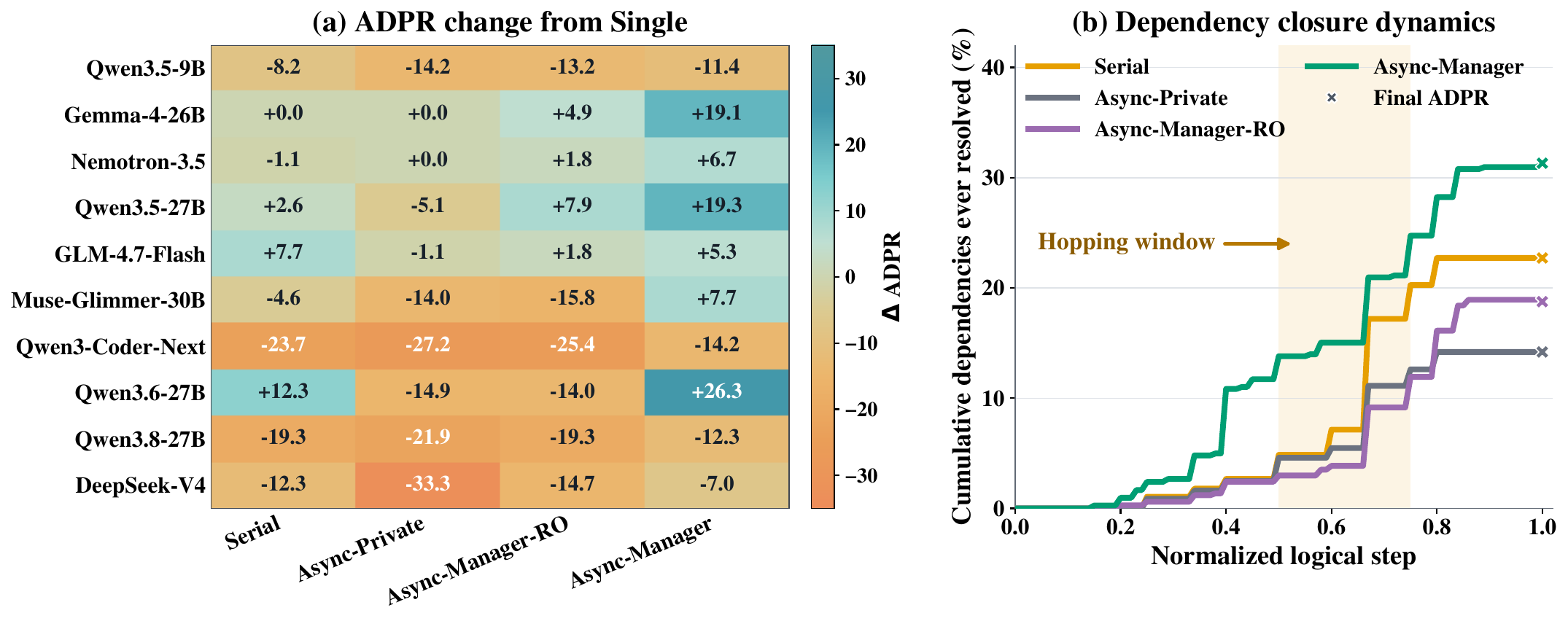}
    \caption{Protocol effects and dependency-closure dynamics. (a) ADPR change relative to Single across models and multi-agent protocols. (b) Cumulative dependency closure over normalized logical steps \(k/T\), showing a recurrent middle-to-late hopping window (\(0.50 < k/T \leq 0.75\)).}
    \label{fig:protocol-effects}
\end{figure*}

Figure~\ref{fig:protocol-effects}(a) examines how execution protocols affect dependency resolution by comparing each collaborative setting against \textsc{Single} under the same model. The results reveal substantial model-dependent variation: a protocol that improves dependency resolution for one model may degrade it for another, and even the same model can exhibit opposite outcomes across protocols. For example, Qwen3.6-27B ranges from a 14.9\% decrease under \textsc{Async-Private} to a 26.3\% improvement under \textsc{Async-Manager}, yielding a 41.2\% spread. Moreover, no collaborative protocol consistently improves dependency resolution across all models. These results indicate that collaboration gains depend not only on the underlying model but also on how multi-agent execution is organized.

Figure~\ref{fig:protocol-effects}(b) reveals a further pattern that is invisible from final ADPR alone. Using the first-resolution checkpoints captured by DRS, we plot the cumulative fraction of dependencies that have achieved closure over normalized logical progress. The horizontal axis represents each checkpoint's relative position in its execution trajectory, while the vertical axis measures the cumulative fraction of dependencies ever resolved by that point.

Across all four collaborative protocols, the curves exhibit a pronounced increase in the middle-to-late portion of the observed trajectories. Across the aggregated trajectories, a substantial fraction of eventual dependency closures concentrates within \(0.50<x\leq0.75\). We refer to this region as a representative \textbf{\textit{hopping window}}, while noting that the onset and duration of the hopping process vary across individual models and protocols. This interval accounts for 34.9\%--67.8\% of eventual closure mass across protocols. This observation suggests that multi-agent collaboration cannot be fully characterized by its initial decomposition or final outcome. When dependencies become resolved is itself an important dimension of collaboration capability. Appendix~\ref{app:aggregate-hopping} further examines the prevalence of this pattern and the effect of checkpoint normalization.

\begin{figure*}[t]
    \centering
    \includegraphics[width=\textwidth]{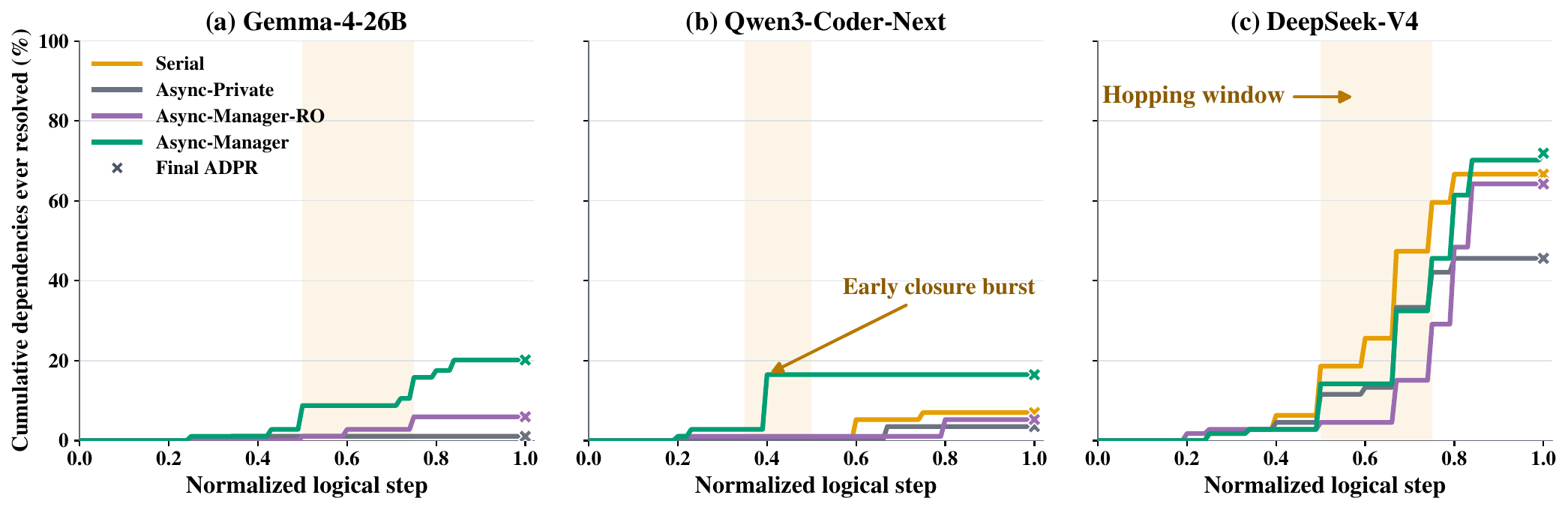}
\caption{Representative dependency-closure trajectories across execution protocols. Each panel shows the cumulative fraction of dependencies ever resolved over normalized logical steps for one model. The shaded region marks the hopping window.}
    \label{fig:rq3-closure-cases}
\end{figure*}

The aggregate trajectories in Figure~\ref{fig:protocol-effects}(b) suggest that the hopping process is closely tied to the formation of substantial dependency resolution. Figure~\ref{fig:rq3-closure-cases} makes this relationship more explicit at the model level. For Gemma-4-26B, protocols whose trajectories remain nearly flat establish little dependency closure, whereas \textsc{Async-Manager} develops a pronounced hopping phase and reaches a substantially higher final ADPR. Qwen3-Coder-Next shows that the location of this phase is not fixed: its \textsc{Async-Manager} trajectory undergoes a sharp closure burst much earlier in execution, around \(x\approx0.4\), before settling into its final plateau. DeepSeek-V4 exhibits another form of the same phenomenon, with all four protocols undergoing pronounced bursts of dependency closure at later stages of their trajectories.

Across these cases, the common factor is not a particular temporal interval, but the emergence of a concentrated phase in which a substantial share of dependencies becomes resolved. This pattern also holds across the full set of 40 model--protocol configurations: 28 exhibit closure growth within the representative interval \(0.50 < x \leq 0.75\), including all 16 configurations reaching at least \(20\%\) final ADPR, while configurations with little final dependency resolution often remain nearly flat throughout execution. The onset and duration of the hopping process nevertheless vary substantially across models and protocols, as illustrated by the earlier burst of Qwen3-Coder-Next. These results suggest that substantial dependency resolution is often established through a concentrated hopping process rather than steady accumulation, while the location of that process is configuration-dependent. Complete trajectories for all models are provided in Appendix~\ref{app:model-level-hopping}.

\subsection{Quality--Cost Trade-offs}
\label{exp:quality_pareto}

Beyond collaboration effectiveness, we examine the token costs incurred by different execution protocols. Multi-agent execution introduces additional computational overhead, yet greater token consumption does not necessarily translate into better dependency resolution. To examine this trade-off, we construct within-model Pareto frontiers, identifying protocols that offer quality gains not attainable at a lower token cost. We use mean reported tokens per task as the cost measure and independently evaluate quality using Test Pass Rate and ADPR.

\begin{wrapfigure}{r}{0.50\textwidth}
\centering
\includegraphics[width=\linewidth]{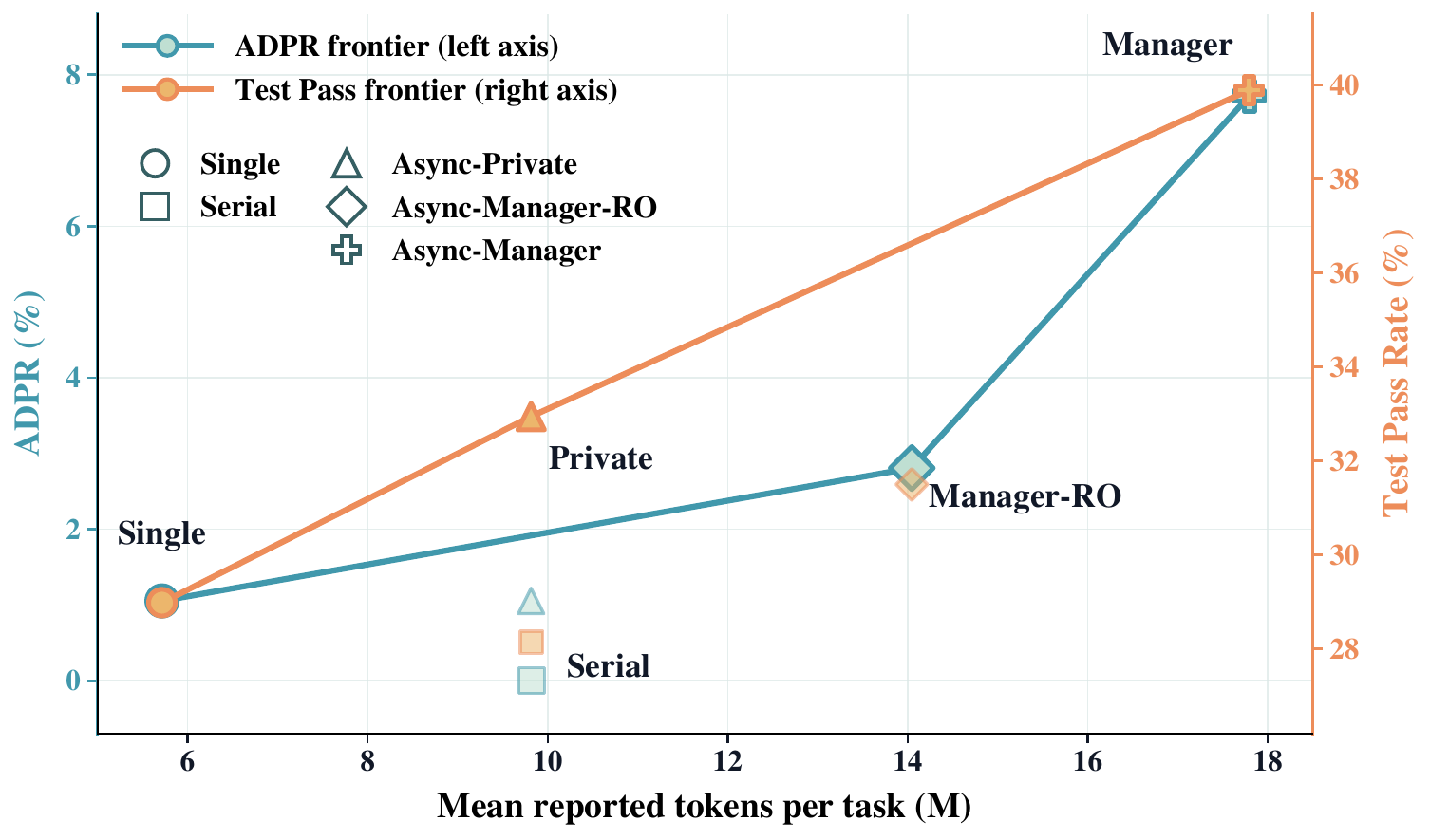}
\caption{Quality--cost Pareto frontiers for Nemotron-3.5 under \textcolor{adprcolor}{ADPR} and \textcolor{tprcolor}{Test Pass Rate}. Both use identical protocols and token costs, yet identify different non-dominated configurations. Marker shapes indicate execution protocols, and lines connect their respective Pareto frontiers.}
\label{fig:quality-cost}
\end{wrapfigure}

Figure~\ref{fig:quality-cost} illustrates this metric-dependent trade-off using Nemotron-3.5. Compared with Single, Async-Private consumes more tokens and improves Test Pass Rate from 28.99\% to 32.94\%, but achieves the same ADPR of 1.05\%. It therefore lies on
the Test Pass Rate frontier but is dominated by Single under ADPR.
Conversely, Async-Manager-RO improves ADPR to 2.81\%, placing it on the ADPR frontier. It is dominated by Async-Private under Test Pass Rate, as it incurs higher token costs while achieving lower test performance.

These differences reveal that the quality--cost assessment of collaboration depends on the metric used to measure quality. Across all ten models, five exhibit different within-model Pareto frontiers under the two metrics, as detailed in Appendix~\ref{app:quality-cost}. Extending our earlier observations of metric mismatch, these results show that task-level performance alone may overlook dependency-resolution gains when evaluating the resource efficiency of multi-agent collaboration. ADPR provides a complementary, dependency-centric perspective on these trade-offs.

%% file: content/Discussion.tex

\section{Conclusion}
Asynchronous multi-agent software engineering introduces a challenge beyond task completion: independently produced components must be coordinated correctly. AsynCodeBench studies this challenge through repository-grounded dependencies, making collaboration and its evolution directly measurable during execution. Our results reveal a persistent gap between standalone coding capability and collaborative performance, while also showing that conventional task-level outcomes can hide unresolved coordination failures. By exposing these hidden dependencies and their dynamics over execution, AsynCodeBench provides a complementary view of multi-agent software engineering beyond final task success. We hope this benchmark supports the development of coding agents that are effective both individually and collaboratively.

\section*{AI Use Statement}

We used OpenAI ChatGPT to assist with language polishing and improving the clarity of the manuscript, literature retrieval and discovery, brainstorming and refinement of research ideas, providing feedback on research methodology and experimental design, and code development and debugging. All methodological choices, benchmark construction, experimental configurations, result verification, and scientific conclusions were independently reviewed, determined, and validated by the authors. The authors take full responsibility for the content of this paper.

%% file: content/Appendix.tex
\newpage
\section{Task Details}

\subsection{Task-Level Dependency Graphs}
\label{subsec:Depedency_Task}

For each benchmark task, we first identify the repository-grounded implementation units that must be modified or completed to satisfy the task specification. We then construct a directed graph
\[
\mathcal{G}=(\mathcal{V},\mathcal{A}),
\]
where \(\mathcal{V}\) contains the implementation units and \(\mathcal{A}\) contains directed producer--consumer dependencies among them. An arc \(d=(u,v)\in\mathcal{A}\) is included only when the correctness of consumer \(v\) depends on behavior, state, an interface, or another artifact provided by producer \(u\). The direction therefore follows the flow of required behavior from producer to consumer.

Dependency arcs are grounded in repository evidence rather than inferred from file proximity or task decomposition alone. For each candidate relation, we inspect the relevant source code, referenced symbols or interfaces, task decomposition, and available public tests to verify that the consumer's correctness genuinely depends on the producer. Relations are excluded when the two implementation units merely co-occur in the same task but can be implemented correctly without satisfying a cross-unit dependency.

The resulting graphs are finalized before model evaluation and remain fixed across all models and execution protocols. Each dependency is manually reviewed for both its direction and supporting evidence, and disagreements during construction are resolved through additional inspection of the repository and associated tests. Freezing the graph before evaluation prevents model behavior or observed outcomes from influencing which dependencies are included.

\begin{figure}[h]
    \centering
    \includegraphics[width=\linewidth]{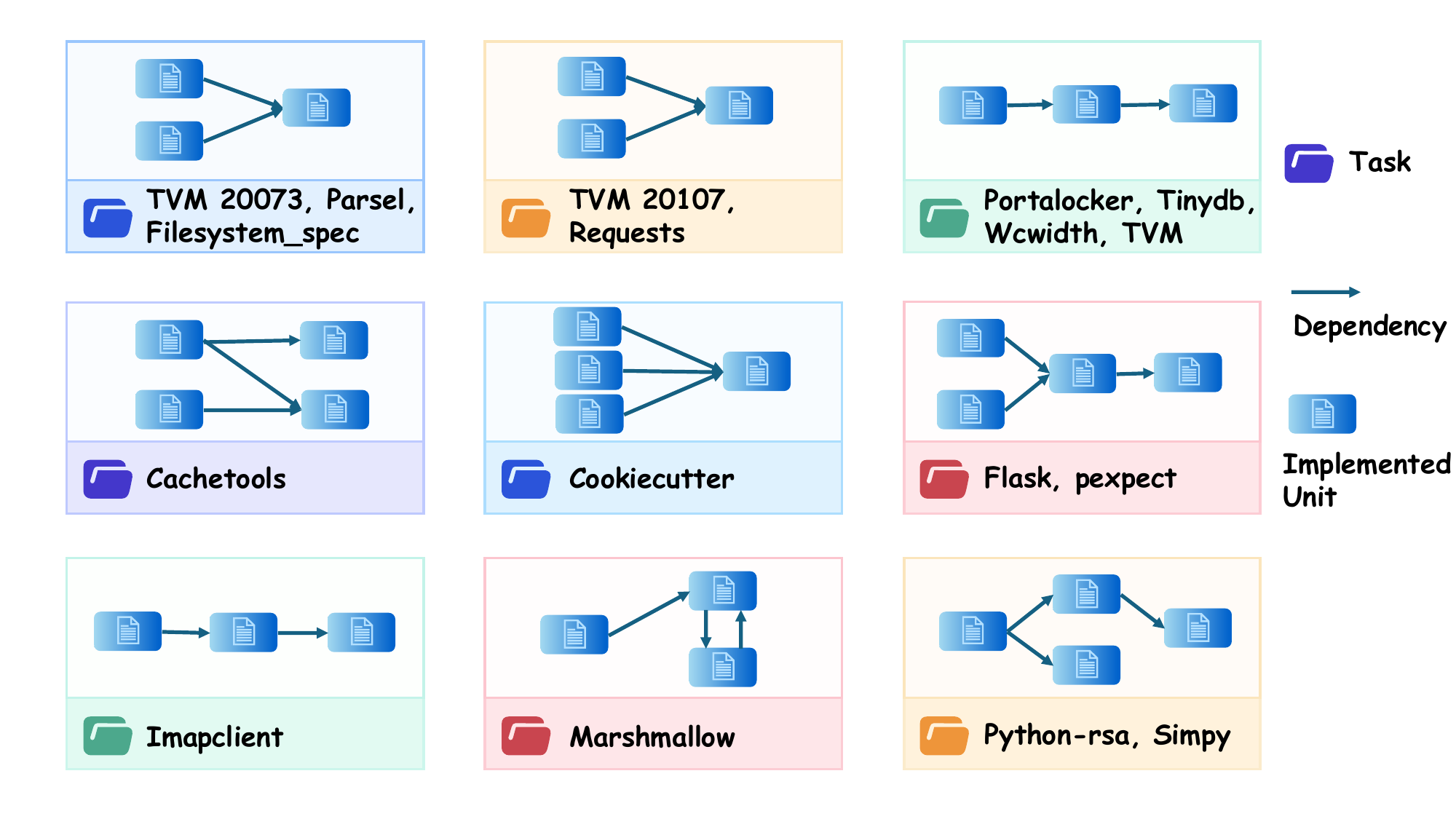}
    \caption{Task-level dependency graphs in AsynCodeBench. Across the benchmark, we identify nine types of dependency relations, each defined by explicit upstream and downstream implementation units. These dependencies form the basis for dependency-level evaluation with ADPR.}
    \label{fig:dependency_graphs}
\end{figure}

We present the dependency graph structure for each task. As shown in Figure~\ref{fig:dependency_graphs}, AsynCodeBench contains nine types of dependency relations. Each dependency specifies well-defined upstream and downstream implementation units, providing the basis for dependency-level evaluation with ADPR.

\subsection{Task Inventory and Provenance}
\label{app:task-inventory}

AsynCodeBench contains 19 tasks from 16 public repositories, covering 64 repository-grounded implementation units and 52 directed dependency edges. Fifteen tasks are derived from
\begin{wrapfigure}{r}{0.4\linewidth}
    \centering
    \includegraphics[width=\linewidth]{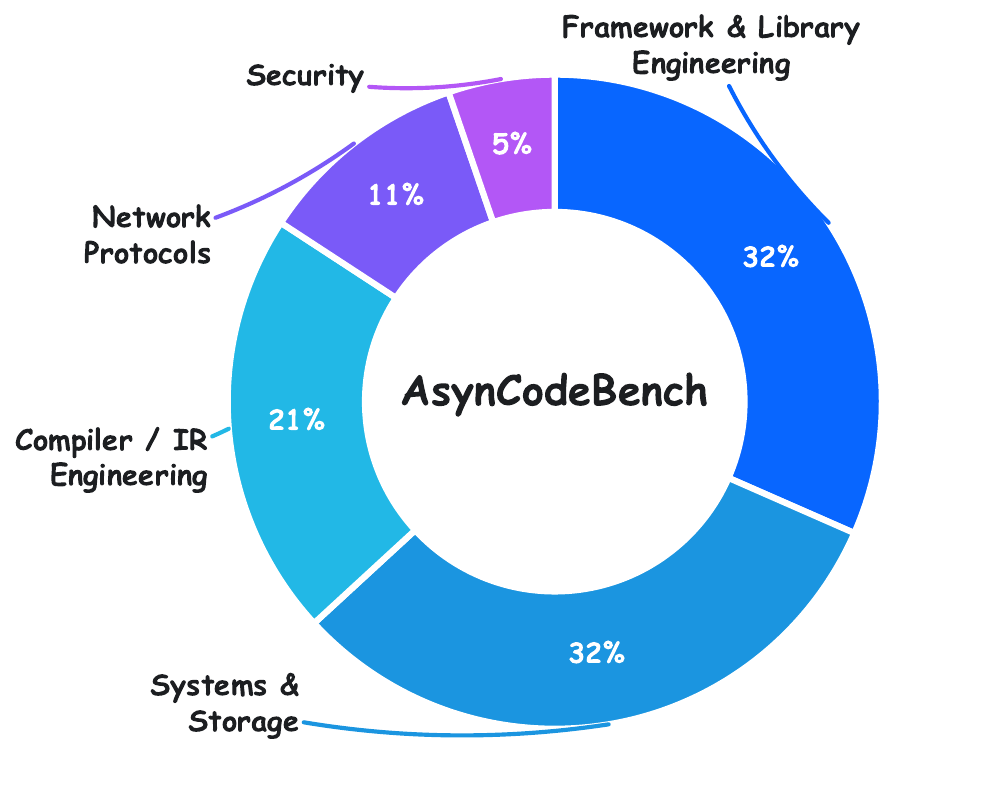}
    \caption{Domain distribution of tasks in AsynCodeBench.}
    \label{fig:distribution}
\end{wrapfigure}Commit0~\citep{zhao2024commit0librarygenerationscratch} and four from merged Apache TVM pull requests; 16 tasks are Python-only, while three involve both Python and C++. Evaluating every task under the five matched execution protocols produces 95 task--protocol scenarios.

Across all task packages, the evaluators reference 148 task-local public test files and collect 4,884 \texttt{pytest} items in completed-state sanity runs. The 52 dependency edges are instrumented with upstream, downstream, and integrated checker groups, yielding 156 checker groups and 508 selector references in total: 125 upstream, 162 downstream, and 221 integrated. After within-task de-duplication, these correspond to 274 task-qualified unique test selectors. Task-level evaluators measure overall implementation correctness, whereas Dependency Checkers provide edge-level observations over the execution trajectory; individual selectors may therefore be shared across multiple edges or checker groups.

\begin{table}[t]
\centering
\caption{Complete AsynCodeBench task inventory and provenance. 
\emph{Agents} denotes the number of specialists used in a multi-agent condition; 
\emph{Files} denotes the number of distinct implementation files writable by specialists; 
\emph{Edges} denotes the number of registered dependency edges; 
\emph{Tests} denotes the number of test files included in the final evaluator; 
and \emph{Ov.} denotes the number of checksum-pinned public overlays. 
Domain abbreviations are F/L (framework and libraries), S/S (systems and storage), 
C/IR (compiler and IR), N/P (network protocols), and SEC (security).}
\label{tab:task-inventory}

\setlength{\tabcolsep}{4pt}

\begin{tabular}{@{}lllcrrrrr@{}}
\toprule
Task & Source record & Domain & Lang. & Agents & Files & Edges & Tests & Ov. \\
\midrule
cachetools       & Commit0         & F/L  & Py     & 2 &  2 & 5 & 12 &  0 \\
deprecated       & Commit0         & F/L  & Py     & 2 &  2 & 3 &  8 &  0 \\
portalocker      & Commit0         & S/S  & Py     & 2 &  2 & 3 &  5 &  1 \\
tinydb           & Commit0         & S/S  & Py     & 2 &  7 & 3 &  8 &  1 \\
wcwidth          & Commit0         & S/S  & Py     & 2 &  2 & 2 &  4 &  0 \\
requests         & Commit0         & N/P  & Py     & 3 & 13 & 3 &  5 &  3 \\
simpy            & Commit0         & S/S  & Py     & 4 &  9 & 3 & 10 &  0 \\
parsel           & Commit0         & F/L  & Py     & 3 &  5 & 2 &  6 &  1 \\
filesystem\_spec & Commit0         & S/S  & Py     & 3 &  5 & 2 &  4 &  4 \\
marshmallow      & Commit0         & F/L  & Py     & 3 &  9 & 3 & 12 &  4 \\
imapclient       & Commit0         & N/P  & Py     & 3 &  9 & 3 &  9 &  5 \\
pexpect          & Commit0         & S/S  & Py     & 4 &  9 & 3 &  4 &  1 \\
flask            & Commit0         & F/L  & Py     & 4 & 10 & 3 &  7 & 11 \\
python-rsa       & Commit0         & SEC  & Py     & 4 &  9 & 3 &  8 &  0 \\
cookiecutter     & Commit0         & F/L  & Py     & 4 & 15 & 3 & 22 &  0 \\
apache-tvm-20018 & TVM PR \#20018 & C/IR & Py+C++ & 3 & 28 & 2 & 14 &  1 \\
apache-tvm-20073 & TVM PR \#20073 & C/IR & Py+C++ & 3 &  7 & 2 &  2 &  1 \\
apache-tvm-20107 & TVM PR \#20107 & C/IR & Py+C++ & 3 & 22 & 2 &  5 &  1 \\
apache-tvm-20153 & TVM PR \#20153 & C/IR & Py     & 3 &  6 & 2 &  3 &  2 \\
\bottomrule
\end{tabular}
\end{table}

\subsection{Human Review Details}
\label{app:human_review_details}

Three reviewers independently examined each benchmark task to verify that it contains meaningful inter-component dependencies and admits a natural decomposition across agents. Each reviewer assessed three fields: \texttt{include}, indicating whether the task satisfies the inclusion criteria of AsynCodeBench; \texttt{parallelizability\_label}, categorized as \emph{parallelizable}, \emph{partially parallelizable}, or \emph{effectively serial}; and \texttt{rationale}, providing the justification for the decision. For each dependency edge, reviewers also verified its validity and direction, together with the supporting repository evidence from source code, symbols, tests, or executable Dependency Checkers; disagreements were resolved through discussion. Reviewer judgments were additionally compared against the benchmark dependency graph, and any discrepancy was explicitly recorded. Table~\ref{tab:human-review-sheet} provides an example review sheet illustrating the human-review process used for benchmark validation. The results are in our anonymized repository.

\begin{table}[t]
\centering
\caption{Human-review sheet used for task validation. Each task is reviewed
independently by three reviewers.}
\label{tab:human-review-sheet}

\begingroup
\setlength{\tabcolsep}{6pt}
\renewcommand{\arraystretch}{1.22}

\begin{tabularx}{\textwidth}{
    >{\raggedright\arraybackslash}p{0.22\textwidth}
    >{\raggedright\arraybackslash}X
}
\toprule
\rowcolor{tablehead}
\multicolumn{2}{l}{\textbf{Task Review Sheet}} \\
\midrule

\textbf{Task}
& \texttt{<task-name>} \\

\textbf{Reviewer}
& Reviewer 1 \\

\midrule

\textbf{Include}
& $\square$ Yes
\qquad
$\square$ No \\

\textbf{Parallelizability}
&
$\square$ Parallelizable
\qquad
$\square$ Partially parallelizable
\qquad
$\square$ Effectively serial \\

\textbf{Rationale}
&
Reviewer justification for whether the task contains meaningful dependencies
and can be naturally decomposed across agents. \\

\bottomrule
\end{tabularx}
\endgroup
\end{table}

\section{Protocol Details}
\label{app:protocl_details}
This section provides the full specification of the five execution protocols summarized in Table~\ref{tab:protocol-details}. Across all protocols, we hold fixed the initial repository state, task instruction, model configuration, final evaluator, and Dependency Checkers. The four multi-agent protocols additionally share the same manifest-defined specialist roles and writable paths, and all specialist artifacts are scope-validated before integration. The protocols differ only in how specialist work is scheduled, what intermediate state is visible, whether a manager coordinates execution, and whether the manager may submit scoped production-code patches.

\begin{table}[t]
\centering
\caption{Execution protocols in AsynCodeBench. ``Writable manager'' denotes scope-constrained patches produced in a private manager worktree; the manager never directly modifies the integrated workspace.}
\label{tab:protocol-details}
\begin{tabular}{lcccc}
\toprule
Protocol & Decomp. & Concurrent & Manager & Writable Manager \\
\midrule
\textsc{Single}           & --         & --         & --         & -- \\
\textsc{Serial}           & \checkmark & --         & --         & -- \\
\textsc{Async-Private}    & \checkmark & \checkmark & --         & -- \\
\textsc{Async-Manager-RO} & \checkmark & \checkmark & \checkmark & -- \\
\textsc{Async-Manager}    & \checkmark & \checkmark & \checkmark & \checkmark \\
\bottomrule
\end{tabular}
\end{table}

\paragraph{\textsc{Single}.}
\textsc{Single} assigns the complete task to one agent with access to the full repository and task instruction. The agent edits the task workspace directly and produces the final implementation without specialist decomposition, handoffs, or manager intervention. This protocol serves as the full-task ownership baseline, avoiding cross-agent coordination while also forgoing specialization and concurrent execution. Because only the final integrated workspace is observed, \textsc{Single} is included in Test Pass Rate and ADPR comparisons but excluded from trajectory-based metrics such as DRS.

\paragraph{\textsc{Serial}.}
\textsc{Serial} uses the manifest-defined specialist decomposition but executes specialists sequentially in dependency order. A downstream specialist begins only after the relevant upstream artifact has been scope-validated and integrated, and therefore observes the latest integrated repository state when starting its work. This protocol preserves explicit specialist handoffs while minimizing stale-context effects, at the cost of concurrency.

\paragraph{\textsc{Async-Private}.}
\textsc{Async-Private} uses the same specialist roles and writable scopes as \textsc{Serial}, but runs specialists concurrently in isolated private worktrees created from the same initial repository state. Specialists cannot observe one another's in-flight changes or receive adaptive feedback during their initial execution. Completed artifacts are scope-validated and integrated in a deterministic dependency order. The contrast with \textsc{Serial} therefore isolates the effect of asynchronous execution under the same task decomposition: dependent implementations are produced independently and must compose only after integration.

\paragraph{\textsc{Async-Manager-RO}.}
\textsc{Async-Manager-RO} retains concurrent private specialist execution and introduces a read-only manager. The manager may inspect the task specification, dependency graph, repository state, specialist artifacts, merge outcomes, and checker diagnostics; it may also plan work, provide feedback, request scoped follow-up attempts, and coordinate integration. It cannot modify production code, so every production patch must originate from a scope-constrained specialist.

This separation follows the orchestrator--worker pattern used in prior multi-agent systems~\citep{fourney2024magenticone,anthropic2024effectiveagents}, where a coordinating agent delegates work to specialized agents; related planner-mediated coordination has also been explored in domain-specific multi-agent systems~\citep{shi2026emr}. Role-specific tool permissions provide a further mechanism for separating coordination from execution~\citep{anthropic2026subagents}. A related design appears in Anthropic's reported DeepSearchQA configuration, where the orchestrator acts through delegated subagents rather than direct tools~\citep{anthropic2026opus46systemcard}. We use read-only authority as an experimental control: it allows explicit coordination without turning the manager into an additional full-context coding agent, while keeping production changes attributable to specialists.

\paragraph{\textsc{Async-Manager}.}
\textsc{Async-Manager} preserves the same coordination workflow but additionally permits the manager to submit scoped production-code interventions after specialist integration. The manager operates in its own private worktree, and its writable region is restricted to the union of manifest-defined specialist production paths. Tests, Dependency Checkers, evaluators, task manifests, Git metadata, and out-of-scope files remain protected.

The manager never edits the integrated workspace or specialist worktrees directly. Instead, each proposed patch is recorded, scope-validated, checked against its base revision, committed in the manager worktree, and integrated only if all policy checks pass. Each accepted state-changing intervention creates a new integrated-workspace checkpoint; rejected or no-change interventions do not alter the trajectory. This protocol tests whether coordination benefits from bounded direct repair in addition to planning, feedback, and delegation.

These are controlled system-level contrasts, not an assumed capability ordering. More generally, agentic protocols can themselves introduce execution overhead or alter model behavior beyond the underlying model capability~\citep{zhang2026tooltax}, motivating our explicit comparison of execution protocols rather than treating the surrounding agent framework as neutral.

\section{Integrated-Workspace Checkpoint Construction}
\label{app:checkpoint-construction}

\begin{figure}[t]
    \centering
    \includegraphics[width=0.9\columnwidth]{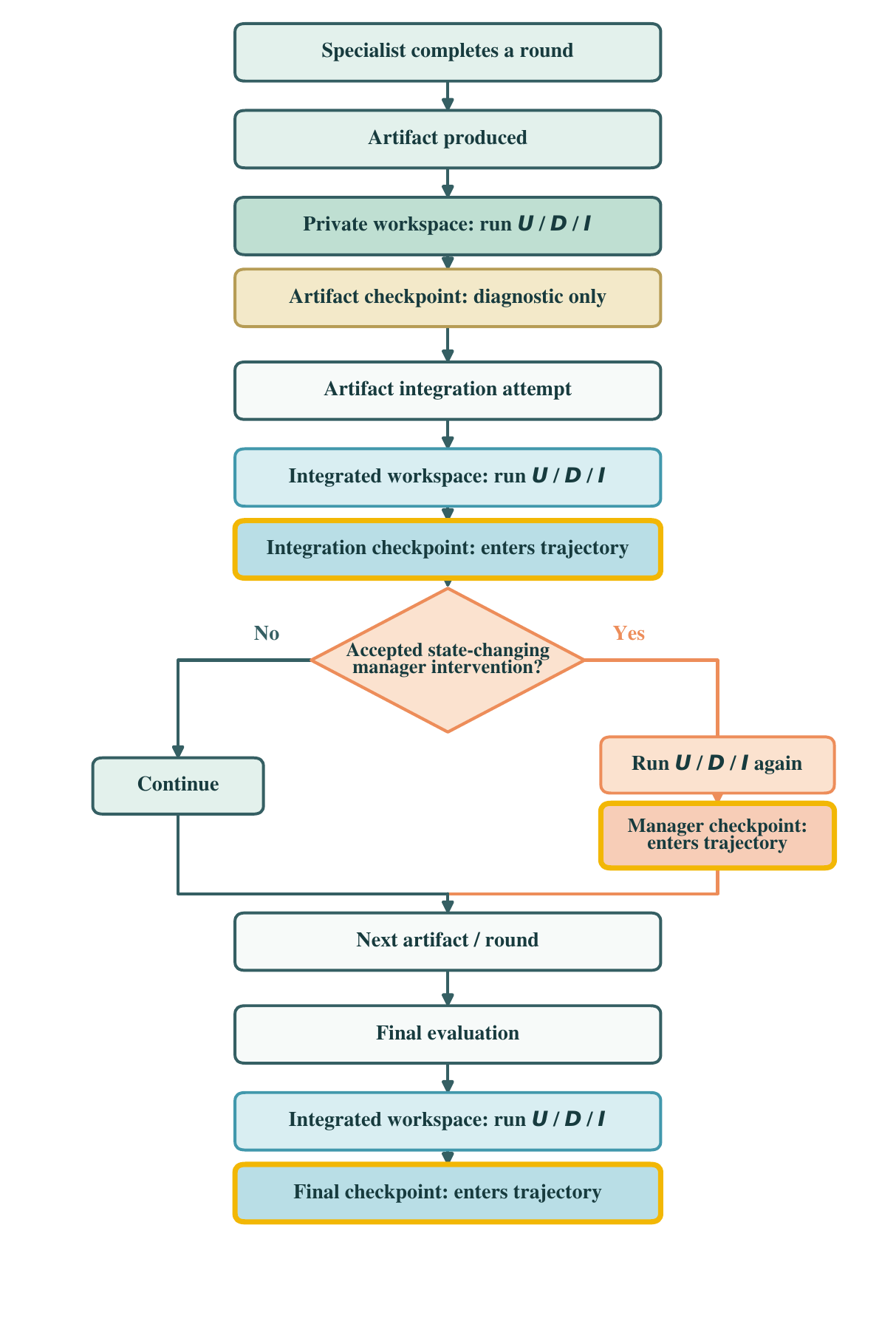}
    \caption{Event-triggered checkpoint construction in AsynCodeBench. The artifact checkpoint evaluates the three checker groups in a private specialist workspace and is used only for local diagnostics. Integrated checker outcomes are recorded at the gold-bordered trajectory checkpoints after integration attempts, accepted state-changing manager interventions when applicable, and final evaluation. Only these ordered integrated-workspace outcomes form the dependency trajectory used by ADPR and DRS.}
    \label{fig:checkpoint-construction}
\end{figure}

Figure~\ref{fig:checkpoint-construction} illustrates how dependency checkpoints are constructed during execution. The harness records checker outcomes in both private specialist workspaces and the integrated workspace. Private artifact checkpoints are used only for local diagnostics and are not included in the dependency trajectory. Instead, we retain the integrated-workspace observations recorded after specialist integration attempts, accepted state-changing manager interventions when applicable, and final evaluation. These retained observations are ordered by execution order and indexed consecutively as \(t=1,\ldots,T\).

Because the number of integration events and accepted manager interventions varies across runs, \(T\) is run-specific. Consequently, the checkpoint index represents logical progress through the observed integrated-workspace trajectory rather than a model turn, tool call, unit of wall-clock time, or fixed amount of computation. \textsc{Single} exposes only the final integrated workspace and is therefore evaluated with ADPR but excluded from temporal DRS analysis.

\section{Evaluated Model Checkpoints}
\label{app:model-details}

Table~\ref{tab:model-details} reports the exact model checkpoints used in our
experiments and their corresponding names in the main paper. Because the model
set includes both dense and mixture-of-experts (MoE) architectures, we report
total and per-token activated parameter counts separately. For dense models,
the activated parameter count equals the language-model parameter count. For
MoE models, the activated count represents the approximate subset of
parameters used for each token and can be substantially smaller than the total
model size. Weight formats such as FP8 and BF16 describe numerical precision
and do not change the parameter count.

\newcommand{\modelicon}[1]{%
    \IfFileExists{fig/logos/#1.png}{%
        \includegraphics[
            width=1.05em,
            height=1.05em,
            keepaspectratio
        ]{fig/logos/#1.png}%
    }{}%
}

\begin{table*}[t]
\centering
\caption{Models evaluated in AsynCodeBench. \emph{Active}
denotes the approximate number of parameters activated per
token. For dense models, all language-model parameters are
active.}
\label{tab:model-details}

\begingroup

\setlength{\tabcolsep}{4pt}

\rowcolors{2}{tablegray}{white}

\begin{tabularx}{\textwidth}{
    l
    >{\hsize=1.28\hsize\linewidth=\hsize
      \raggedright\arraybackslash}X
    >{\hsize=1.68\hsize\linewidth=\hsize
      \raggedright\arraybackslash}X
    >{\hsize=.85\hsize\linewidth=\hsize
      \raggedright\arraybackslash}X
    >{\hsize=.55\hsize\linewidth=\hsize
      \centering\arraybackslash}X
    >{\hsize=.64\hsize\linewidth=\hsize
      \centering\arraybackslash}X
}

\toprule
\rowcolor{tablehead}
\multicolumn{2}{l}{\textbf{Paper name}}
& \textbf{Evaluated checkpoint}
& \makecell[l]{\textbf{Archi-}\\\textbf{tecture}}
& \textbf{Total}
& \textbf{Active} \\
\midrule

\modelicon{qwen}
& \textbf{Qwen3.5-9B}
& \texttt{Qwen/}\par
  \texttt{Qwen3.5-9B}
& Dense & 9B & 9B \\

\modelicon{qwen}
& \textbf{Qwen3.5-27B}
& \texttt{Qwen/}\par
  \texttt{Qwen3.5-27B}
& Dense & 27B & 27B \\

\modelicon{qwen}
& \textbf{Qwen3.6-27B}
& \texttt{Qwen/}\par
  \texttt{Qwen3.6-27B}
& Dense & 27B & 27B \\

\modelicon{qwen}
& \textbf{Qwen3.8-27B}
& \texttt{Qwen/}\par
  \texttt{Qwen3.8-27B}
& Dense & 27B & 27B \\

\modelicon{qwen}
& \textbf{Qwen3-Coder-}\par
  \textbf{Next}
& \texttt{Qwen/}\par
  \texttt{Qwen3-Coder-}\par
  \texttt{Next-FP8}
& MoE & 80B & 3B \\

\modelicon{deepseek}
& \textbf{DeepSeek-}\par
  \textbf{V4-Flash}
& \texttt{deepseek/}\par
  \texttt{deepseek-v4-}\par
  \texttt{flash-0731}
& MoE & 284B & 13B \\

\modelicon{google}
& \textbf{Gemma-4-26B}
& \texttt{google/}\par
  \texttt{gemma-4-26B-}\par
  \texttt{A4B-it}
& MoE & 25.2B & 3.8B \\

\modelicon{meta}
& \textbf{Muse-Glimmer-}\par
  \textbf{30B}
& \texttt{meta-models/}\par
  \texttt{Muse-Glimmer-}\par
  \texttt{30B}
& Dense
& ${\sim}$29.6B
& ${\sim}$29.6B \\

\modelicon{GLM}
& \textbf{GLM-4.7-Flash}
& \texttt{zai-org/}\par
  \texttt{GLM-4.7-Flash}
& MoE & 30B & 3B \\

\modelicon{nvidia}
& \textbf{Nemotron-}\par
  \textbf{3.5-Lightning}
& \texttt{nvidia/}\par
  \texttt{NVIDIA-Nemotron-}\par
  \texttt{3.5-Lightning-}\par
  \texttt{30B-A3B-BF16}
& \makecell[l]{Hybrid\\MoE}
& 30B & 3B \\

\bottomrule

\end{tabularx}

\endgroup
\end{table*}

\section{Test Progress and Dependency Resolution Across Protocols}
\label{appendix:test_pass_rate_adpr}

\begin{figure*}[ht]
    \centering
    \includegraphics[width=\textwidth]{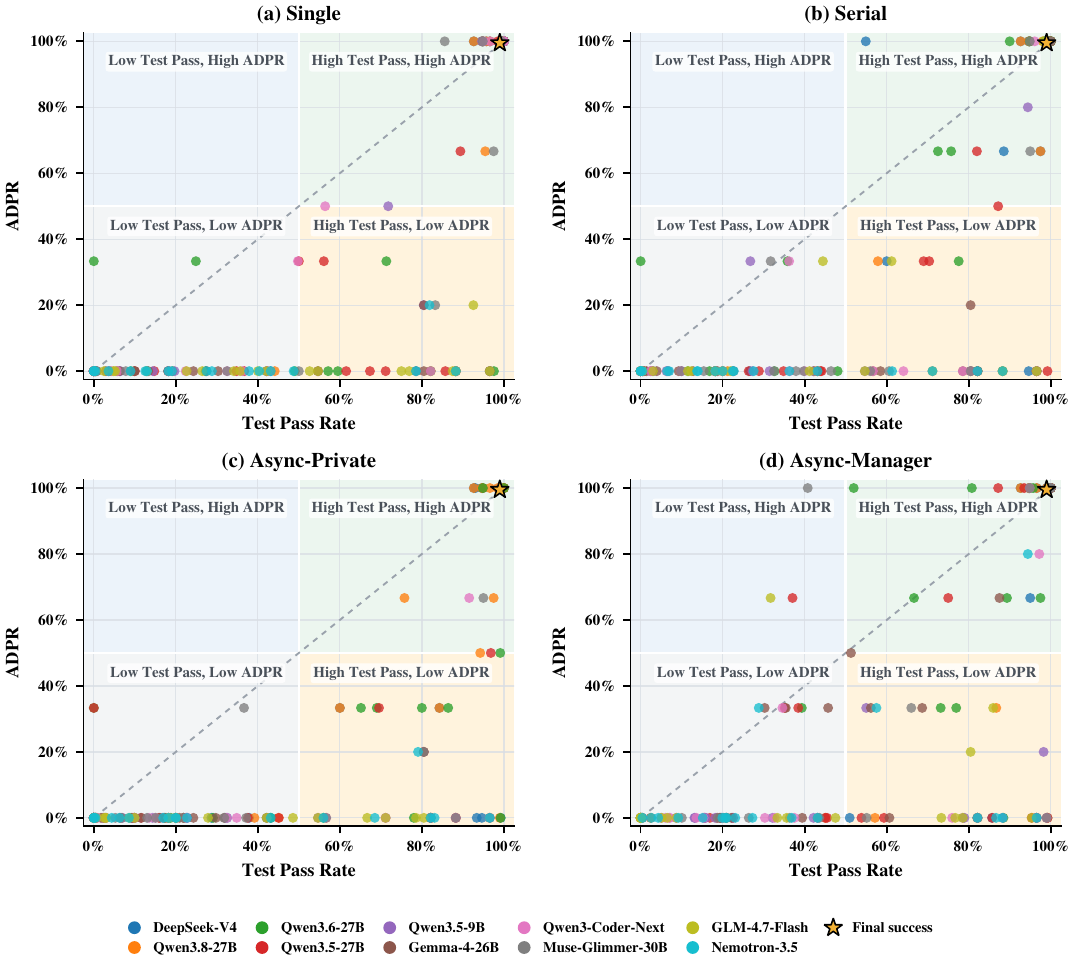}
    \caption{Test Pass Rate versus ADPR under the four protocols not shown in Figure~\ref{fig:collaboration-gap}(b). Each point represents a model--task outcome and colors denote models. The dashed line indicates equality between Test Pass Rate and ADPR, while the 50\% boundaries provide a common visual partition. All panels use identical axes and encodings.}
    \label{fig:app-protocol-pass-adpr}
\end{figure*}

Section~\ref{sec:collaboration-gap} uses \textsc{Async-Manager-RO} to illustrate the mismatch between task-level test progress and dependency resolution. Figure~\ref{fig:app-protocol-pass-adpr} extends the same analysis to the remaining four protocols. Each panel contains 190 model--task outcomes from the same ten models and 19 tasks, with identical axes and visual encodings to enable direct comparison.

The same asymmetric mismatch appears across protocols. As shown in Table~\ref{tab:app-protocol-regions}, high-test, low-ADPR outcomes account for 22.1--27.4\% of observations under every protocol. By contrast, the low-test, high-ADPR region is empty for four protocols and contains only three outcomes under \textsc{Async-Manager}. Thus, substantial test progress frequently occurs without corresponding dependency closure, whereas the reverse pattern is rare.

\begin{table*}[t]
\centering
\caption{Distribution of model--task outcomes across the four regions in Figure~\ref{fig:app-protocol-pass-adpr} and Figure~\ref{fig:collaboration-gap}(b). Each protocol contains 190 outcomes. Entries report count (percentage). The 50\% threshold is used only as a descriptive visual partition. Blue cells denote the best value according to the indicated direction, while orange cells highlight values discussed in the accompanying analysis.}
\label{tab:app-protocol-regions}

\rowcolors{2}{tablegray}{white}
\begin{tabularx}{\textwidth}{lYYYY}
\toprule
\rowcolor{tablehead}
\textbf{Protocol} &
\makecell{\textbf{Low Test /}\\ \textbf{Low ADPR} $\downarrow$} &
\makecell{\textbf{Low Test /}\\ \textbf{High ADPR}} &
\makecell{\textbf{High Test /}\\ \textbf{Low ADPR} $\downarrow$} &
\makecell{\textbf{High Test /}\\ \textbf{High ADPR} $\uparrow$} \\
\midrule
\textsc{Single}
& 97 (51.1\%)
& 0 (0.0\%)
& \best{42 (22.1\%)}
& 51 (26.8\%) \\

\textsc{Serial}
& 104 (54.7\%)
& 0 (0.0\%)
& 44 (23.2\%)
& 42 (22.1\%) \\

\textsc{Async-Private}
& 116 (61.1\%)
& 0 (0.0\%)
& 49 (25.8\%)
& 25 (13.2\%) \\

\textsc{Async-Manager-RO}
& 105 (55.3\%)
& 0 (0.0\%)
& \focus{52 (27.4\%)}
& 33 (17.4\%) \\

\textsc{Async-Manager}
& \best{83 (43.7\%)}
& \focus{3 (1.6\%)}
& 50 (26.3\%)
& \best{54 (28.4\%)} \\
\bottomrule
\end{tabularx}
\end{table*}

Table~\ref{tab:app-protocol-decoupling} provides complementary continuous diagnostics. Test Pass Rate and ADPR remain positively associated under all five protocols, with Pearson correlations of 0.599--0.716 and Spearman correlations of 0.559--0.737. However, Test Pass Rate exceeds ADPR in 62.6--77.9\% of outcomes, with mean signed gaps of 23.7--29.2\%. Moreover, 12.6--16.8\% of outcomes achieve at least 80\% Test Pass Rate while resolving fewer than half of their declared dependencies. The discrepancy observed in Section~\ref{sec:collaboration-gap} is therefore not specific to \textsc{Async-Manager-RO}.

\begin{table*}[t]
\centering
\caption{Correlation and mismatch diagnostics by protocol. Percentages are computed over 190 model--task outcomes per protocol. The signed gap is Test Pass Rate minus ADPR; the absolute gap ignores direction. Arrows indicate the preferred direction where applicable. \protect\colorbox{bestblue}{Blue} cells denote the best value according to the indicated direction, while  \protect\colorbox{focusorange}{orange} cells highlight values discussed in the accompanying analysis.}
\label{tab:app-protocol-decoupling}

\rowcolors{2}{tablegray}{white}
\begin{tabularx}{\textwidth}{lYYYYY}
\toprule
\rowcolor{tablehead}
\textbf{Metric}
& \textbf{Single}
& \textbf{Serial}
& \textbf{Async-P.}
& \textbf{Manager-RO}
& \textbf{Manager} \\
\midrule

Pearson $r$
& 0.716
& 0.667
& 0.599
& \focus{0.617}
& 0.628 \\

Spearman $\rho$
& 0.737
& 0.651
& 0.559
& \focus{0.613}
& 0.657 \\

Mean Test Pass Rate $\uparrow$
& 51.0\%
& 47.9\%
& 41.9\%
& \focus{48.0\%}
& \best{57.4\%} \\

Mean ADPR $\uparrow$
& 27.4\%
& 22.7\%
& 14.2\%
& \focus{18.8\%}
& \best{31.3\%} \\

Mean signed gap $\downarrow$
& \best{23.7\%}
& 25.2\%
& 27.8\%
& \focus{29.2\%}
& 26.1\% \\

Test Pass Rate $>$ ADPR $\downarrow$
& \best{62.6\%}
& 70.5\%
& 77.9\%
& \focus{71.6\%}
& 67.4\% \\

Test Pass $\geq80\%$, ADPR $<50\%$ $\downarrow$
& \best{12.6\%}
& \best{12.6\%}
& 15.8\%
& \focus{16.8\%}
& 16.3\% \\

Test Pass $\geq80\%$, ADPR $=0$ $\downarrow$
& \best{10.0\%}
& 12.1\%
& 12.1\%
& \focus{13.7\%}
& 14.2\% \\

Final task success $\uparrow$
& \best{21.6\%}
& 13.2\%
& 8.4\%
& 13.2\%
& 21.1\% \\

\bottomrule
\end{tabularx}
\end{table*}

\section{Diagnosing Dependency Closure Dynamics}
\label{app:closure-dynamics}

\subsection{Aggregate Closure Concentration and Normalization}
\label{app:aggregate-hopping}

Figure~\ref{fig:protocol-effects}(b) shows that first dependency closures are not distributed uniformly over the aggregated execution trajectories. A particularly strong concentration appears within \(0.50 < x \leq 0.75\), where \(x\) denotes the normalized logical step. We treat this interval as a representative region of the aggregate hopping phenomenon rather than as a fixed temporal definition, since the onset and duration of concentrated closure may vary across individual models and protocols.

Table~\ref{tab:hopping-window} quantifies this aggregate concentration. The first three columns report the cumulative fraction of dependencies that have achieved closure by the halfway point, the three-quarter point, and at least once during execution, respectively. The final column reports the fraction of eventual closure mass accumulated within the representative interval. Under \textsc{Serial}, for example, cumulative closure increases from 4.9\% to 20.3\% between \(x=0.50\) and \(x=0.75\), accounting for 67.8\% of its eventual closure mass. Across the four protocols, 34.9\%--67.8\% of eventual closure mass is accumulated within this interval, confirming that a substantial portion of dependency resolution is concentrated within a relatively short segment of the execution trajectory.

\begin{table*}[t]
\centering
\caption{Cumulative dependency closure and hopping-window concentration across execution protocols.}
\label{tab:hopping-window}
\rowcolors{3}{tablegray}{white}
\begin{tabular}{lcccc}
\toprule
\rowcolor{tablehead}
& \multicolumn{3}{c}{Cumulative closure (\%)} & \\
\cmidrule(lr){2-4}
\rowcolor{tablehead}
Protocol & By 50\% & By 75\% & Ever closed & Closure in $(0.50,0.75]$ (\%) \\
\midrule
\textsc{Serial}           & 4.9  & 20.3 & 22.7 & 67.8 \\
\textsc{Async-Private}    & 4.6  & 12.6 & 14.2 & 56.5 \\
\textsc{Async-Manager-RO} & 3.0  & 11.9 & 18.9 & 47.3 \\
\textsc{Async-Manager}    & 13.8 & 24.7 & 31.3 & 34.9 \\
\bottomrule
\end{tabular}
\end{table*}

A natural concern is whether this concentration is induced by the normalization used to align trajectories of different lengths. Let \(k\) denote the first integrated checkpoint at which a dependency closes and \(T\) the total number of integrated checkpoints in the corresponding run, giving the normalized position \(x=k/T\). Although this mapping allows trajectories with different numbers of checkpoints to be compared on a common scale, discrete checkpoint structures can accumulate at shared ratios such as \(2/3\).

Table~\ref{tab:closure-diagnostic} shows that normalization indeed sharpens the aggregate peak: 50.0\%--73.9\% of first closures around this region correspond to \(k=2,T=3\). However, the underlying concentration remains visible in raw checkpoint space. Across all four protocols, \(k=2\) is still the most common first-closure checkpoint, accounting for 38.4\%--60.2\% of eventual closures. Normalization therefore accentuates the observed hopping pattern but does not create the underlying burst of dependency closure.

\begin{table*}[t]
\centering
\caption{Effect of checkpoint normalization on the hopping-window pattern.}
\label{tab:closure-diagnostic}
\rowcolors{2}{tablegray}{white}
\begin{tabular}{lcc}
\toprule
\rowcolor{tablehead}
Protocol & Peak events from \(k=2,T=3\) & First closure at raw \(k=2\) \\
\midrule
\textsc{Serial}           & 68.9\% & 58.6\% \\
\textsc{Async-Private}    & 73.9\% & 60.2\% \\
\textsc{Async-Manager-RO} & 66.7\% & 38.4\% \\
\textsc{Async-Manager}    & 50.0\% & 48.6\% \\
\bottomrule
\end{tabular}
\end{table*}

\begin{figure*}[t]
    \centering
    \includegraphics[width=0.88\textwidth]{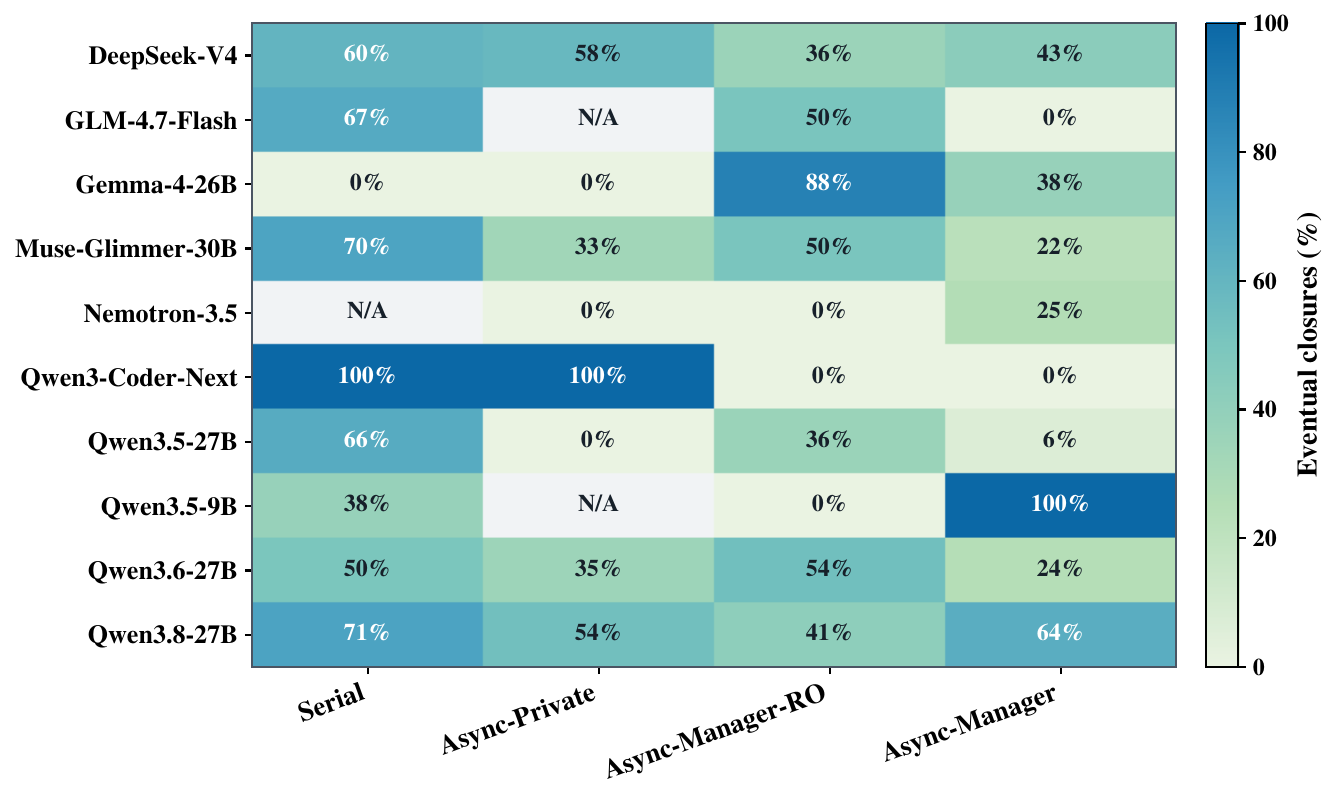}
    \caption{Model--protocol variation in hopping-window concentration. Each cell shows the share of eventual first closures occurring within the hopping window (0.50, 0.75]. N/A denotes no observed dependency closure.}
    \label{fig:hopping-window-heatmap}
\end{figure*}

The aggregate concentration, however, does not imply that every model--protocol configuration follows the same temporal pattern. Figure~\ref{fig:hopping-window-heatmap} exposes substantial heterogeneity beneath the protocol-level averages: while some configurations place a large share of their eventual first closures within \(0.50<x\leq0.75\), others contribute little or no closure in this region. The interval therefore captures where closure is most strongly concentrated in aggregate, but not a fixed boundary that governs every execution. This distinction motivates a closer examination of the complete model-level trajectories in Section~\ref{app:model-level-hopping}, where we show that the hopping process can emerge earlier or later depending on the model and execution protocol.

\subsection{Model-Level Variation in Hopping Dynamics}
\label{app:model-level-hopping}

\begin{figure*}[t]
    \centering
    \includegraphics[width=\textwidth]{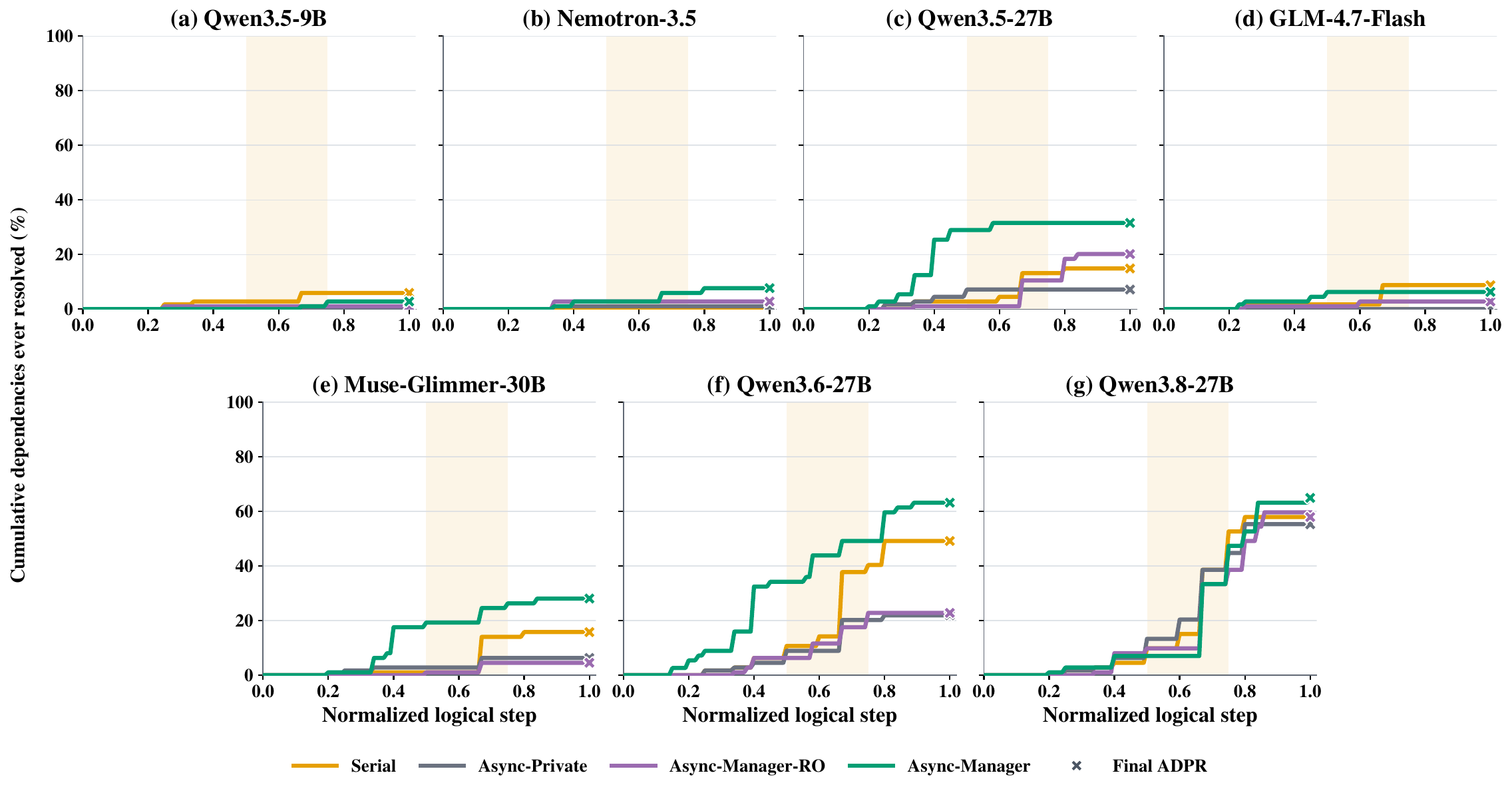}
    \caption{Dependency-closure trajectories for the seven models not shown in the main-text case studies. Curves report cumulative dependencies ever resolved over normalized logical execution, and crosses denote final ADPR. The shaded region marks the representative aggregate concentration \(0.50<x\leq0.75\). Individual trajectories show that the onset and magnitude of concentrated closure vary substantially across models and protocols.}
    \label{fig:app-closure-dynamics}
\end{figure*}

The variation in Figure~\ref{fig:hopping-window-heatmap} raises a natural question: if the aggregate hopping concentration is not shared uniformly across configurations, how does dependency closure actually unfold at the model level? Figure~\ref{fig:app-closure-dynamics} shows the complete trajectories for the remaining seven models, complementing the three representative cases in Figure~\ref{fig:rq3-closure-cases}.

The trajectories reveal that the main variation lies not in whether closure must occur within a particular normalized interval, but in when and how strongly a concentrated closure phase emerges. Low-closure configurations, such as many protocols for Qwen3.5-9B, Nemotron-3.5, and GLM-4.7-Flash, remain largely flat throughout execution. By contrast, configurations with substantial final closure exhibit pronounced bursts, but their timing differs considerably. For Qwen3.5-27B and Qwen3.6-27B, \textsc{Async-Manager} begins accumulating dependencies well before the representative aggregate interval, whereas Qwen3.8-27B concentrates much of its closure later in the trajectory. Similar differences also appear across protocols within the same model. These trajectories therefore support a broader interpretation of hopping as a configuration-dependent phase of concentrated dependency closure, whose onset and magnitude vary across models and execution protocols.

\section{Additional Quality--Cost Analysis}
\label{app:quality-cost}

To complement the Nemotron-3.5 case study in the main text, we extend the quality--cost analysis to all ten evaluated models. For each model, we independently construct Pareto frontiers using ADPR and Test Pass Rate as alternative quality measures, while keeping the candidate protocols and their token costs fixed. A protocol is dominated if another protocol achieves equal or higher quality at equal or lower token cost, with at least one strict improvement. Table~\ref{tab:app-frontier-comparison} summarizes the resulting differences in frontier membership. Five models exhibit different frontiers, yielding two ADPR-only and four Test-Pass-only configurations.

\begin{figure*}[t]
\centering
\includegraphics[width=\textwidth]{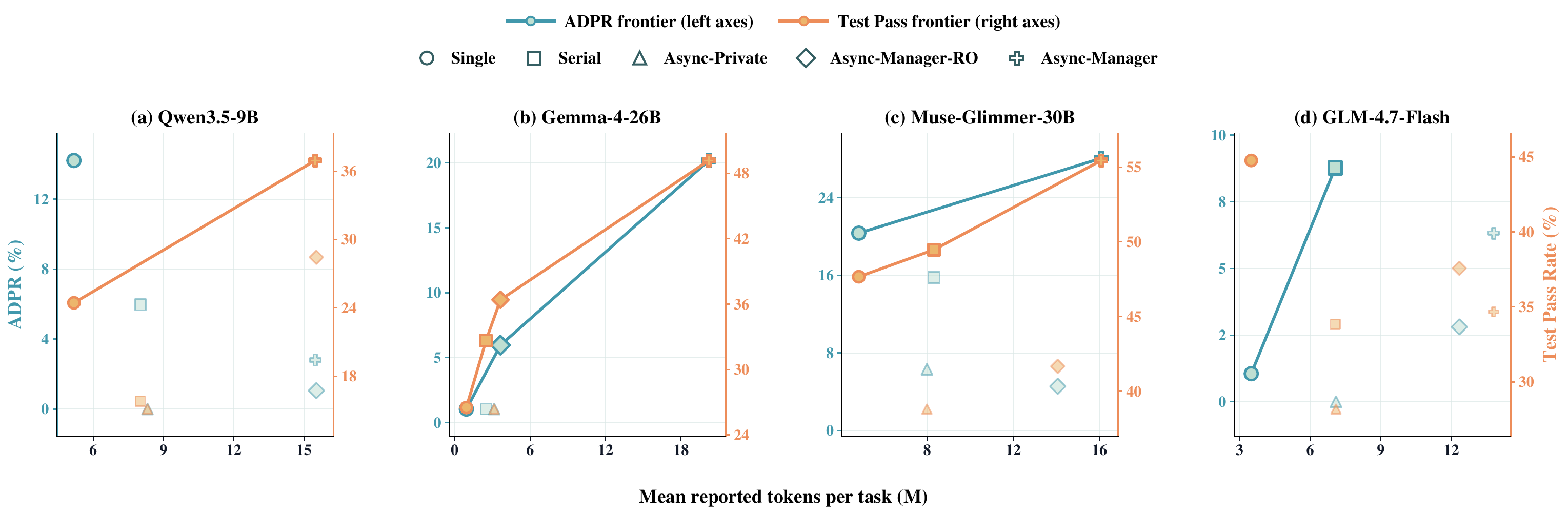}
\caption{Metric-dependent quality--cost trade-offs for four additional models. Each panel shows the within-model Pareto frontiers under ADPR (blue, left axis) and Test Pass Rate (orange, right axis). Marker shapes distinguish execution protocols, and lines connect non-dominated configurations under each metric. Dominated configurations remain visible. Axis ranges are independently scaled across models.}
\label{fig:app-quality-cost}
\end{figure*}

Figure~\ref{fig:app-quality-cost} presents the other four models with differing frontiers. For Qwen3.5-9B, Single achieves the highest ADPR while consuming the fewest tokens, leaving all collaborative protocols dominated under ADPR. However, Async-Manager substantially improves Test Pass Rate and therefore enters the corresponding frontier. A similar metric-dependent pattern appears in Gemma-4-26B and Muse-Glimmer-30B: Serial improves Test Pass Rate over Single but achieves equal or lower ADPR at a higher token cost, making it non-dominated only under Test Pass Rate. GLM-4.7-Flash exhibits the opposite pattern. Serial improves ADPR over Single and enters the ADPR frontier, whereas Single achieves higher Test Pass Rate at a lower token cost, dominating all collaborative protocols under the conventional metric.

These cases demonstrate that improvements in test performance do not necessarily translate into dependency-resolution gains, and vice versa. Consequently, the same execution protocols can yield different non-dominated configurations depending on the quality metric used. Together with the Nemotron-3.5 case study, these results extend the metric mismatch observed in the main experiments to cost-aware evaluation, illustrating the complementary role of dependency-centric metrics. The complete numerical results underlying these comparisons are reported in Tables~\ref{tab:full-results-qwen} and~\ref{tab:full-results-other}.

\begin{table*}[t]
\centering
\caption{Comparison of within-model Pareto frontiers
under ADPR and Test Pass Rate. The two quality metrics
are evaluated using identical protocols and token costs.
Only metric-exclusive configurations are listed.}
\label{tab:app-frontier-comparison}

\rowcolors{2}{tablegray}{white}
\begin{tabularx}{\textwidth}{l c L L}
\toprule
\rowcolor{tablehead}
\textbf{Model}
& \textbf{Same membership?}
& \textbf{ADPR-only protocols}
& \textbf{Test-Pass-only protocols} \\
\midrule

DeepSeek-V4
& Yes & -- & -- \\

Qwen3.8-27B
& Yes & -- & -- \\

Qwen3.6-27B
& Yes & -- & -- \\

Qwen3.5-27B
& Yes & -- & -- \\

Qwen3-Coder-Next
& Yes & -- & -- \\

Qwen3.5-9B
& No & -- & Async-Manager \\

Gemma-4-26B
& No & -- & Serial \\

Muse-Glimmer-30B
& No & -- & Serial \\

GLM-4.7-Flash
& No & Serial & -- \\

Nemotron-3.5
& No & Async-Manager-RO & Async-Private \\

\bottomrule
\end{tabularx}
\end{table*}

\section{Case Study: When High Test Pass Rate Masks Unresolved Dependencies}
\label{app:tvm-20073-case}

\begin{figure*}[t]
    \centering
    \includegraphics[width=\textwidth]{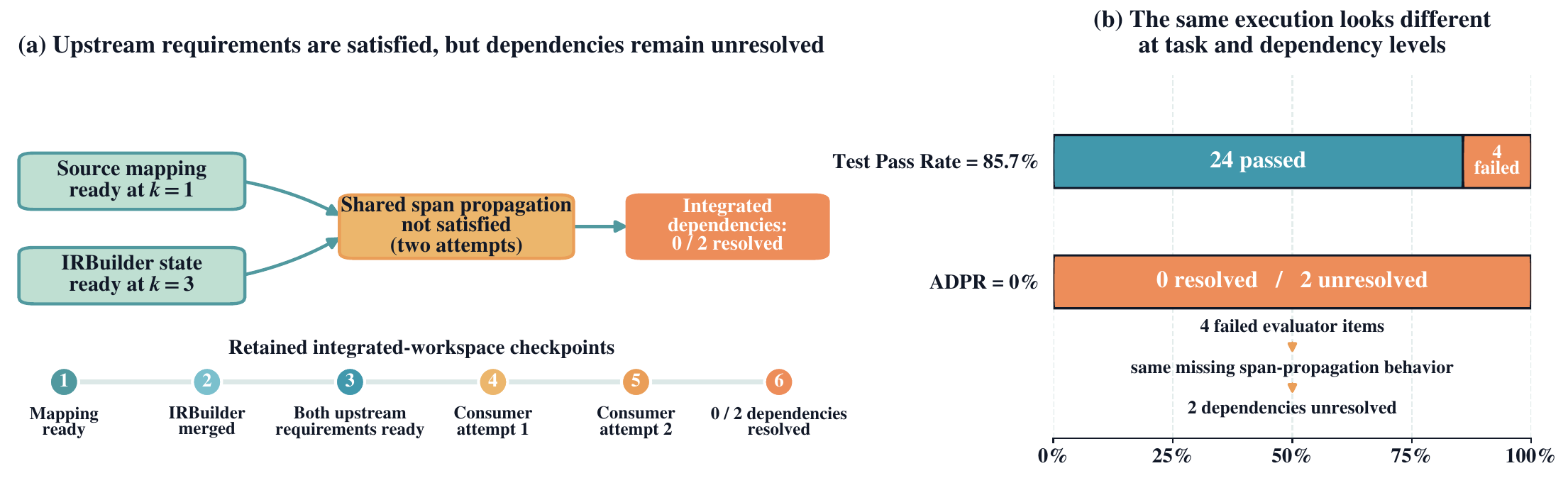}
    \caption{
    Case study of DeepSeek-V4 under \textsc{Async-Manager} on \textsc{TVM-20073}. 
    Although both upstream requirements become available during execution, the shared downstream span-propagation logic remains unsatisfied after two attempts, leaving both dependencies unresolved. 
    The final workspace nevertheless passes 24 of 28 evaluator items (85.7\% Test Pass Rate), while ADPR remains 0\%, illustrating how task-level performance can obscure unresolved cross-agent dependencies and the coordination failures behind them.
    }
    \label{fig:tvm-20073-case}
\end{figure*}

Figure~\ref{fig:tvm-20073-case} illustrates why task-level and dependency-level evaluation can lead to different views of the same execution. Under DeepSeek-V4 with \textsc{Async-Manager} on \textsc{TVM-20073}, the final workspace passes 24 of 28 evaluator items, yielding a Test Pass Rate of 85.7\%. From the final test outcome alone, the implementation appears largely complete.

The dependency trajectory reveals a different structure behind these results. The task contains two dependencies: source-coordinate mapping and IRBuilder active-span state must both be incorporated into a shared span-propagation implementation. Source mapping is available from the first retained checkpoint, and both upstream requirements are satisfied by checkpoint 3. However, the downstream propagation logic remains unsatisfied after two subsequent attempts, leaving both dependencies unresolved at final evaluation and yielding an ADPR of \(0/2=0\%\).

The discrepancy arises because the two metrics count different units. The four failed evaluator items all exercise the same missing span-propagation behavior, whereas ADPR identifies the two unresolved dependencies underlying those failures. Thus, several test failures that appear separately in the final evaluator trace back to a shared coordination bottleneck: the required upstream capabilities are available, but they are never successfully composed in the downstream implementation.

This case highlights the complementary information provided by dependency-aware evaluation. Test Pass Rate summarizes repository-level correctness over evaluator items, while the dependency trajectory identifies whether required cross-agent dependencies are established and when they become available. The latter therefore exposes coordination structure that is not visible from the final test outcome alone.

\section{Detailed Results of All Models}
\label{app:detailed_results}

Tables~\ref{tab:full-results-qwen} and~\ref{tab:full-results-other} report the complete experimental results for all ten models under five execution protocols, covering Test Pass Rate, ADPR, normalized DRS, and mean token consumption per task. The results are aggregated over the same 19 benchmark tasks for each model--protocol configuration, providing the numerical basis for the model capability, protocol, and quality--cost analyses in the main text. Test Pass Rate and ADPR are reported as percentages, with higher values indicating better performance, whereas lower DRS indicates earlier dependency resolution. DRS is not reported for Single, which has no intermediate integration checkpoints. Token consumption is expressed in millions per task; the marked Qwen3.5-9B Async-Manager result represents a lower bound due to one incomplete token record.

We also report the Final Task Success Rate in Table~\ref{tab:task-success-counts}. Unlike the Test Pass Rate, the Final Task Success Rate adopts a stricter criterion: a task is considered successful only if it achieves both a complete task pass and full dependency resolution (ADPR = 1). For each model–protocol combination, we report the average success rate across the 19 tasks.

\begin{table*}[t]
\centering
\caption{Complete experimental results for the Qwen model family.
Test Pass Rate and ADPR are reported as percentages.
Tokens denote mean reported tokens per task in millions.}
\label{tab:full-results-qwen}

\rowcolors{2}{tablegray}{white}
\begin{tabularx}{\textwidth}{lYYYY}
\toprule
\rowcolor{tablehead}
\textbf{Protocol}
& \textbf{Test Pass $\uparrow$}
& \textbf{ADPR $\uparrow$}
& \textbf{DRS $\downarrow$}
& \textbf{Tokens (M) $\downarrow$} \\
\midrule

\rowcolor{bestblue}
\multicolumn{5}{c}{
    \cellcolor{modelband}
    \IfFileExists{fig/logos/qwen.png}{
        \raisebox{-0.16\height}{
            \includegraphics[
                height=1.15em,
                keepaspectratio
            ]{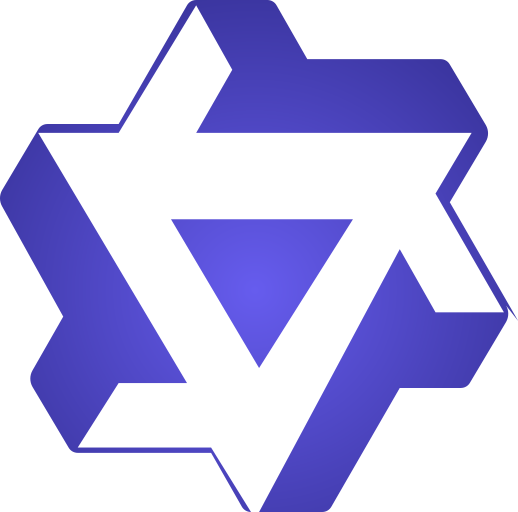}
        }
        \hspace{0.45em}
    }{}
    \textbf{Qwen3.5-9B}
} \\

Single           & 24.44 & 14.21 & --    & 5.18 \\
Serial           & 15.83 & 5.96  & 0.962 & 8.01 \\
Async-Private    & 15.13 & 0.00  & 1.000 & 8.32 \\
Async-Manager-RO & 28.44 & 1.05  & 0.992 & 15.53 \\
Async-Manager    & 36.93 & 2.81  & 0.988 & $15.48^{\dagger}$ \\

\addlinespace[4pt]

\rowcolor{bestblue}
\multicolumn{5}{c}{
    \cellcolor{modelband}
    \IfFileExists{fig/logos/qwen.png}{
        \raisebox{-0.16\height}{
            \includegraphics[
                height=1.15em,
                keepaspectratio
            ]{fig/logos/qwen.png}
        }
        \hspace{0.45em}
    }{}
    \textbf{Qwen3.5-27B}
} \\

Single           & 42.13 & 12.28 & --    & 4.13 \\
Serial           & 47.91 & 14.91 & 0.920 & 6.18 \\
Async-Private    & 31.15 & 7.19  & 0.951 & 6.67 \\
Async-Manager-RO & 49.33 & 20.18 & 0.913 & 12.41 \\
Async-Manager    & 59.15 & 31.58 & 0.787 & 14.00 \\

\addlinespace[4pt]

\rowcolor{bestblue}
\multicolumn{5}{c}{
    \cellcolor{modelband}
    \IfFileExists{fig/logos/qwen.png}{
        \raisebox{-0.16\height}{
            \includegraphics[
                height=1.15em,
                keepaspectratio
            ]{fig/logos/qwen.png}
        }
        \hspace{0.45em}
    }{}
    \textbf{Qwen3.6-27B}
} \\

Single           & 57.45 & 36.84 & --    & 4.96 \\
Serial           & 64.01 & 49.12 & 0.754 & 8.35 \\
Async-Private    & 55.79 & 21.93 & 0.878 & 8.18 \\
Async-Manager-RO & 63.32 & 22.81 & 0.880 & 13.88 \\
Async-Manager    & 76.70 & 63.16 & 0.641 & 14.49 \\

\addlinespace[4pt]

\rowcolor{bestblue}
\multicolumn{5}{c}{
    \cellcolor{modelband}
    \IfFileExists{fig/logos/qwen.png}{
        \raisebox{-0.16\height}{
            \includegraphics[
                height=1.15em,
                keepaspectratio
            ]{fig/logos/qwen.png}
        }
        \hspace{0.45em}
    }{}
    \textbf{Qwen3.8-27B}
} \\

Single           & 93.62 & 77.19 & --    & 3.45 \\
Serial           & 79.53 & 57.89 & 0.717 & 8.02 \\
Async-Private    & 81.76 & 55.26 & 0.727 & 7.09 \\
Async-Manager-RO & 82.33 & 57.89 & 0.733 & 12.78 \\
Async-Manager    & 86.67 & 64.91 & 0.719 & 13.73 \\

\addlinespace[4pt]

\rowcolor{bestblue}
\multicolumn{5}{c}{
    \cellcolor{modelband}
    \IfFileExists{fig/logos/qwen.png}{
        \raisebox{-0.16\height}{
            \includegraphics[
                height=1.15em,
                keepaspectratio
            ]{fig/logos/qwen.png}
        }
        \hspace{0.45em}
    }{}
    \textbf{Qwen3-Coder-Next}
} \\

Single           & 51.18 & 30.70 & --    & 4.69 \\
Serial           & 39.77 & 7.02  & 0.967 & 9.49 \\
Async-Private    & 33.55 & 3.51  & 0.982 & 8.66 \\
Async-Manager-RO & 28.83 & 5.26  & 0.977 & 17.89 \\
Async-Manager    & 45.97 & 16.49 & 0.886 & 16.38 \\

\bottomrule
\end{tabularx}
\end{table*}

\begin{table*}[t]
\centering
\caption{Complete experimental results for the remaining model families.
Test Pass Rate and ADPR are reported as percentages.
Tokens denote mean reported tokens per task in millions.}
\label{tab:full-results-other}

\rowcolors{2}{tablegray}{white}
\begin{tabularx}{\textwidth}{lYYYY}
\toprule
\rowcolor{tablehead}
\textbf{Protocol}
& \textbf{Test Pass $\uparrow$}
& \textbf{ADPR $\uparrow$}
& \textbf{DRS $\downarrow$}
& \textbf{Tokens (M) $\downarrow$} \\
\midrule

\rowcolor{bestblue}
\multicolumn{5}{c}{
    \cellcolor{modelband}
    \IfFileExists{fig/logos/deepseek.png}{
        \raisebox{-0.16\height}{
            \includegraphics[
                height=1.15em,
                keepaspectratio
            ]{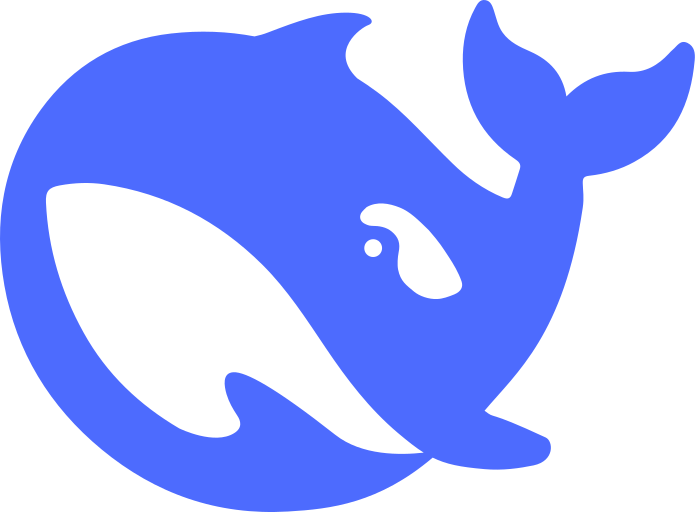}
        }
        \hspace{0.45em}
    }{}
    \textbf{DeepSeek-V4}
} \\

Single           & 93.64 & 78.95 & --    & 2.54 \\
Serial           & 87.84 & 66.67 & 0.665 & 6.89 \\
Async-Private    & 75.75 & 45.61 & 0.769 & 6.92 \\
Async-Manager-RO & 80.36 & 64.21 & 0.748 & 11.61 \\
Async-Manager    & 89.85 & 71.93 & 0.690 & 13.65 \\

\addlinespace[4pt]

\rowcolor{bestblue}
\multicolumn{5}{c}{
    \cellcolor{modelband}
    \IfFileExists{fig/logos/google.png}{
        \raisebox{-0.16\height}{
            \includegraphics[
                height=1.15em,
                keepaspectratio
            ]{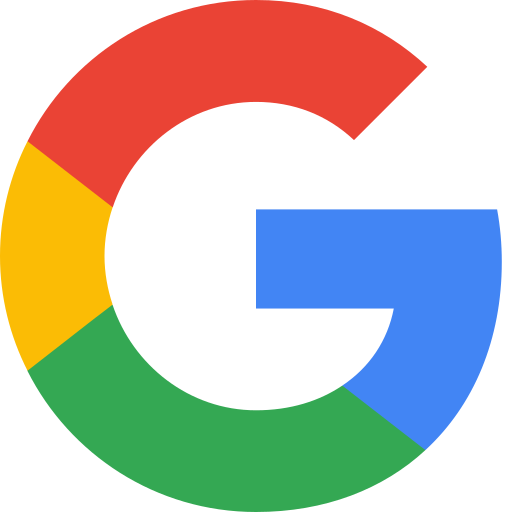}
        }
        \hspace{0.45em}
    }{}
    \textbf{Gemma-4-26B}
} \\

Single           & 26.51 & 1.05  & --    & 0.93 \\
Serial           & 32.65 & 1.05  & 0.992 & 2.49 \\
Async-Private    & 26.37 & 1.05  & 0.992 & 3.14 \\
Async-Manager-RO & 36.40 & 5.96  & 0.972 & 3.64 \\
Async-Manager    & 49.18 & 20.18 & 0.905 & 20.20 \\

\addlinespace[4pt]

\rowcolor{bestblue}
\multicolumn{5}{c}{
    \cellcolor{modelband}
    \IfFileExists{fig/logos/meta.png}{
        \raisebox{-0.16\height}{
            \includegraphics[
                height=1.15em,
                keepaspectratio
            ]{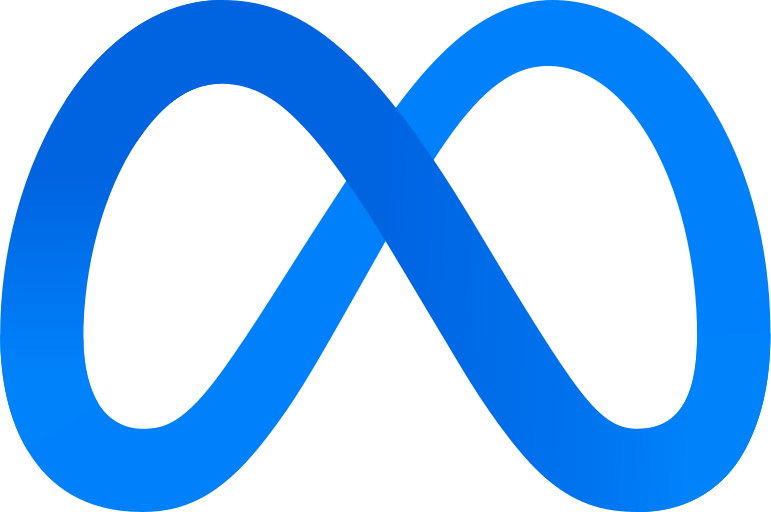}
        }
        \hspace{0.45em}
    }{}
    \textbf{Muse-Glimmer-30B}
} \\

Single           & 47.67 & 20.35 & --    & 4.83 \\
Serial           & 49.48 & 15.79 & 0.921 & 8.32 \\
Async-Private    & 38.81 & 6.32  & 0.961 & 7.99 \\
Async-Manager-RO & 41.67 & 4.56  & 0.976 & 14.06 \\
Async-Manager    & 55.45 & 28.07 & 0.832 & 16.08 \\

\addlinespace[4pt]

\rowcolor{bestblue}
\multicolumn{5}{c}{
    \cellcolor{modelband}
    \IfFileExists{fig/logos/GLM.png}{
        \raisebox{-0.16\height}{
            \includegraphics[
                height=1.15em,
                keepaspectratio
            ]{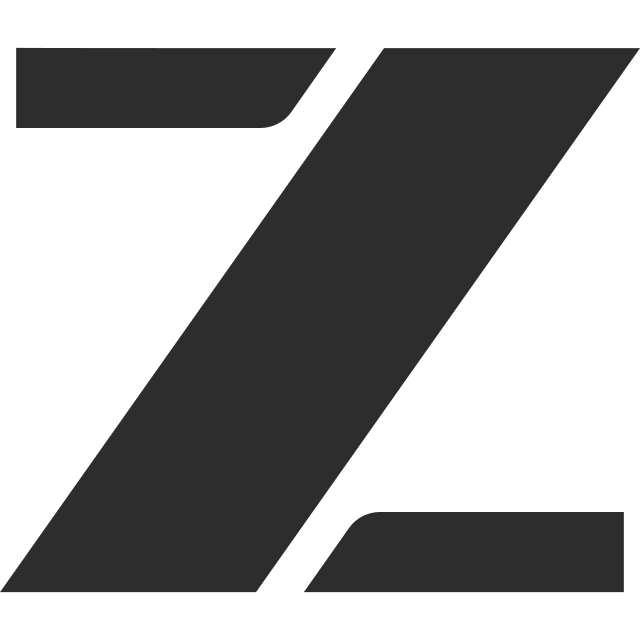}
        }
        \hspace{0.45em}
    }{}
    \textbf{GLM-4.7-Flash}
} \\

Single           & 44.77 & 1.05 & --    & 3.50 \\
Serial           & 33.85 & 8.77 & 0.951 & 7.06 \\
Async-Private    & 28.20 & 0.00 & 1.000 & 7.10 \\
Async-Manager-RO & 37.58 & 2.81 & 0.983 & 12.34 \\
Async-Manager    & 34.68 & 6.32 & 0.957 & 13.79 \\

\addlinespace[4pt]

\rowcolor{bestblue}
\multicolumn{5}{c}{
    \cellcolor{modelband}
    \IfFileExists{fig/logos/nvidia.png}{
        \raisebox{-0.16\height}{
            \includegraphics[
                height=1.15em,
                keepaspectratio
            ]{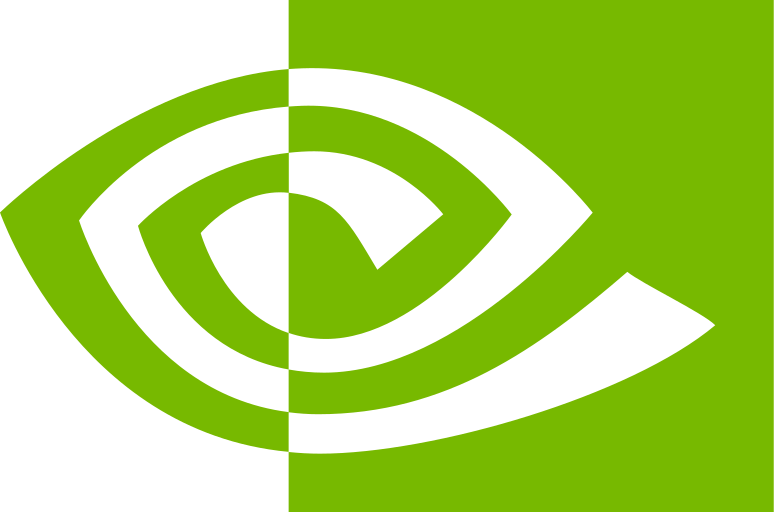}
        }
        \hspace{0.45em}
    }{}
    \textbf{Nemotron-3.5}
} \\

Single           & 28.99 & 1.05 & --    & 5.72 \\
Serial           & 28.14 & 0.00 & 1.000 & 9.82 \\
Async-Private    & 32.94 & 1.05 & 0.992 & 9.82 \\
Async-Manager-RO & 31.50 & 2.81 & 0.980 & 14.04 \\
Async-Manager    & 39.89 & 7.72 & 0.959 & 17.79 \\

\bottomrule
\end{tabularx}
\end{table*}

\definecolor{modelband}{RGB}{244,241,235}

\newcommand{\modelrowstrut}{\rule[-0.52em]{0pt}{2.50em}}
\newcommand{\datarowstrut}{\rule[-0.36em]{0pt}{2.05em}}

\begin{table*}[t]
\centering
\caption{Number of tasks with successful outcomes across models and execution protocols. Each entry reports the number of successful tasks out of 19, with the corresponding percentage in parentheses.}
\label{tab:task-success-counts}

\begin{tabularx}{\textwidth}{YYYYY}
\toprule
\rowcolor{tablehead}
\textbf{Single}
& \textbf{Serial}
& \textbf{Async-Private}
& \textbf{Async-Manager-RO}
& \textbf{Async-Manager} \\
\midrule

\rowcolor{bestblue}
\multicolumn{5}{c}{
    \cellcolor{modelband}
    \modelrowstrut
    \IfFileExists{fig/logos/qwen.png}{
        \raisebox{-0.16\height}{
            \includegraphics[height=1.15em,keepaspectratio]{fig/logos/qwen.png}
        }
        \hspace{0.45em}
    }{}
    \textbf{Qwen3.5-9B}
} \\
\datarowstrut 2/19 (10.5\%)
& 0/19 (0.0\%)
& 0/19 (0.0\%)
& 0/19 (0.0\%)
& 0/19 (0.0\%) \\

\rowcolor{bestblue}
\multicolumn{5}{c}{
    \cellcolor{modelband}
    \modelrowstrut
    \IfFileExists{fig/logos/google.png}{
        \raisebox{-0.16\height}{
            \includegraphics[height=1.15em,keepaspectratio]{fig/logos/google.png}
        }
        \hspace{0.45em}
    }{}
    \textbf{Gemma-4-26B}
} \\
\datarowstrut 0/19 (0.0\%)
& 0/19 (0.0\%)
& 0/19 (0.0\%)
& 0/19 (0.0\%)
& 1/19 (5.3\%) \\

\rowcolor{bestblue}
\multicolumn{5}{c}{
    \cellcolor{modelband}
    \modelrowstrut
    \IfFileExists{fig/logos/nvidia.png}{
        \raisebox{-0.16\height}{
            \includegraphics[height=1.15em,keepaspectratio]{fig/logos/nvidia.png}
        }
        \hspace{0.45em}
    }{}
    \textbf{Nemotron-3.5}
} \\
\datarowstrut 0/19 (0.0\%)
& 0/19 (0.0\%)
& 0/19 (0.0\%)
& 0/19 (0.0\%)
& 0/19 (0.0\%) \\

\rowcolor{bestblue}
\multicolumn{5}{c}{
    \cellcolor{modelband}
    \modelrowstrut
    \IfFileExists{fig/logos/qwen.png}{
        \raisebox{-0.16\height}{
            \includegraphics[height=1.15em,keepaspectratio]{fig/logos/qwen.png}
        }
        \hspace{0.45em}
    }{}
    \textbf{Qwen3.5-27B}
} \\
\datarowstrut 0/19 (0.0\%)
& 1/19 (5.3\%)
& 0/19 (0.0\%)
& 3/19 (15.8\%)
& 2/19 (10.5\%) \\

\rowcolor{bestblue}
\multicolumn{5}{c}{
    \cellcolor{modelband}
    \modelrowstrut
    \IfFileExists{fig/logos/glm.png}{
        \raisebox{-0.16\height}{
            \includegraphics[height=1.15em,keepaspectratio]{fig/logos/glm.png}
        }
        \hspace{0.45em}
    }{}
    \textbf{GLM-4.7-Flash}
} \\
\datarowstrut 0/19 (0.0\%)
& 0/19 (0.0\%)
& 0/19 (0.0\%)
& 0/19 (0.0\%)
& 0/19 (0.0\%) \\

\rowcolor{bestblue}
\multicolumn{5}{c}{
    \cellcolor{modelband}
    \modelrowstrut
    \IfFileExists{fig/logos/meta.png}{
        \raisebox{-0.16\height}{
            \includegraphics[height=1.15em,keepaspectratio]{fig/logos/meta.png}
        }
        \hspace{0.45em}
    }{}
    \textbf{Muse-Glimmer-30B}
} \\
\datarowstrut 1/19 (5.3\%)
& 2/19 (10.5\%)
& 0/19 (0.0\%)
& 0/19 (0.0\%)
& 3/19 (15.8\%) \\

\rowcolor{bestblue}
\multicolumn{5}{c}{
    \cellcolor{modelband}
    \modelrowstrut
    \IfFileExists{fig/logos/qwen.png}{
        \raisebox{-0.16\height}{
            \includegraphics[height=1.15em,keepaspectratio]{fig/logos/qwen.png}
        }
        \hspace{0.45em}
    }{}
    \textbf{Qwen3-Coder-Next}
} \\
\datarowstrut 3/19 (15.8\%)
& 0/19 (0.0\%)
& 0/19 (0.0\%)
& 1/19 (5.3\%)
& 2/19 (10.5\%) \\

\rowcolor{bestblue}
\multicolumn{5}{c}{
    \cellcolor{modelband}
    \modelrowstrut
    \IfFileExists{fig/logos/qwen.png}{
        \raisebox{-0.16\height}{
            \includegraphics[height=1.15em,keepaspectratio]{fig/logos/qwen.png}
        }
        \hspace{0.45em}
    }{}
    \textbf{Qwen3.6-27B}
} \\
\datarowstrut 6/19 (31.6\%)
& 5/19 (26.3\%)
& 2/19 (10.5\%)
& 3/19 (15.8\%)
& 7/19 (36.8\%) \\

\rowcolor{bestblue}
\multicolumn{5}{c}{
    \cellcolor{modelband}
    \modelrowstrut
    \IfFileExists{fig/logos/qwen.png}{
        \raisebox{-0.16\height}{
            \includegraphics[height=1.15em,keepaspectratio]{fig/logos/qwen.png}
        }
        \hspace{0.45em}
    }{}
    \textbf{Qwen3.8-27B}
} \\
\datarowstrut 14/19 (73.7\%)
& 8/19 (42.1\%)
& 7/19 (36.8\%)
& 7/19 (36.8\%)
& 12/19 (63.2\%) \\

\rowcolor{bestblue}
\multicolumn{5}{c}{
    \cellcolor{modelband}
    \modelrowstrut
    \IfFileExists{fig/logos/deepseek.png}{
        \raisebox{-0.16\height}{
            \includegraphics[height=1.15em,keepaspectratio]{fig/logos/deepseek.png}
        }
        \hspace{0.45em}
    }{}
    \textbf{DeepSeek-V4}
} \\
\datarowstrut 15/19 (78.9\%)
& 9/19 (47.4\%)
& 7/19 (36.8\%)
& 11/19 (57.9\%)
& 13/19 (68.4\%) \\

\bottomrule
\end{tabularx}
\end{table*}